\documentclass[a4paper,fleqn,usenatbib]{mnras}
\usepackage{newtxtext,newtxmath}
\usepackage[T1]{fontenc}
\usepackage{ae,aecompl}

\usepackage{graphicx}	% Including figure files
\usepackage{amsmath}	% Advanced maths commands
\usepackage{amssymb}	% Extra maths symbols

\title[Irradiation in XTE J1859$+$226]{Time-Evolved X-ray Irradiation during the 1999--2000 Outburst of the Black-Hole Binary XTE J1859$+$226}

\author[M.~Kimura \& C.~Done]{Mariko Kimura$^{1, 2}$
\thanks{E-mail: mkimura@kusastro.kyoto-u.ac.jp}
and Chris Done$^{2}$
\\
$^{1}$Department of Astronomy, Graduate School of Science, Kyoto University, 
Oiwakecho, Kitashirakawa, Sakyo-ku, Kyoto 606-8502, Japan\\
$^{2}$Department of Physics, University of Durham, South Road, Durham, DH1 3LE, UK\\}

\date{Accepted XXX. Received YYY; in original form ZZZ}

\pubyear{2018}

\begin{document}
\label{firstpage}
\pagerange{\pageref{firstpage}--\pageref{lastpage}}
\maketitle

\begin{abstract}
  \textcolor{black}{
  X-ray irradiation in X-ray binaries is thought to control 
  the behavior at the outer disc, which is observable 
  mainly at optical wavelengths.} 
  It is generally parameterized phenomenologically, 
  \textcolor{black}{but it can also be predicted from theoretical models} 
  of irradiated discs and their coronae/winds. We test these models
  using five multi-wavelength \textit{Hubble Space Telescope (HST)} and
  quasi simultaneous \textit{Rossi X-Ray Timing Explorer (RXTE)} 
  datasets from the black-hole binary XTE J1859$+$226.  
  These show how the reprocessed optical emission changes 
  during outburst as the source fades from the very 
  high/intermediate state at $\sim$0.4$L_{\rm Edd}$ down through 
  the high/soft state towards the transition to the hard state 
  at $\sim$0.02$L_{\rm Edd}$. The models are able to reproduce 
  the small change in reprocessing efficiency as the source flux 
  decreases by a factor of two, and the spectrum softens from 
  the very high/intermediate state to the bright high/soft state. 
  However, the low luminosity high/soft state as well as 
  the transition spectrum show more complex behaviour
  that is not well described by current models.  We suggest 
  the disc geometry has changed drastically during the outburst, 
  probably due to tidal forces, and that the disc is no longer 
  in steady state at the late stage of the outburst. 
\end{abstract}


%%%%%%%%%%%%%%%%%%%%%%%%%%%%%%%%%%%%%%%%%%%%%%%%%%

%%%%%%%%%%%%%%%%% BODY OF PAPER %%%%%%%%%%%%%%%%%%

\section{Introduction}

Low-mass X-ray binaries (LMXBs) have a black hole or 
a neutron star as a central object with a companion 
low mass star which fills its Roche lobe.  Mass transfer 
takes place, forming an accretion disc around the
central object, and this disc is globally unstable 
if it crosses the Hydrogen ionisation temperature 
at any radius.  The disc then cycles between long quiescent 
intervals punctuated by dramatic outbursts, with fast rise 
and exponential decay, during which most of the disc
material is accreted \cite[e.g.,][]{tan96XNreview,che97BHXN}.  
This is rather different to the classic disc instability 
mechanism as first outlined to describe normal dwarf-nova-type 
outbursts, which show rapid decline 
\cite[][for a review]{osa96review}.  
In this kind of outbursts, the H-ionisation triggers 
a heating wave which sweeps across the disc, but the outer 
disc soon dips below the H-ionisation temperature, 
triggering a cooling wave which switches the disc back to
quiescence. The major difference in behaviour between LMXBs 
and dwarf novae is that the central accretion flow is 
much brighter than in dwarf novae due to the higher 
gravitational potential of the central object.  
X-ray irradiation from the inner disc is then important, 
heating the outer disc, and preventing Hydrogen recombination 
from launching the cooling wave which switches off the outburst.  
Such irradiation controlled decays generally 
match well to the observed light curves in LMXBs 
\citep{min89BHXN,kin98SXTlightcurve,kin98ADirradiation,dub01XNmodel,las01DIDNXT,cor12XB}.  

X-ray irradiation of the outer disc should be directly 
observable, as it produces reprocessed optical 
\citep{van94visualLMXB,dej96reprocess}.  
It also produces a hot disc atmosphere/wind 
due to another ionisation instability which operates 
for material in pressure balance (here, hydrostatic 
equilibrium).  X-rays heat the surface, so it expands, 
and its density decreases, and its ionisation state 
increases so that it can be almost completely ionised. 
Deeper down, the material must have larger pressure 
in order to support the weight of the upper layers, 
and then, its density must increase.  Bremsstrahlung 
cooling becomes more important, reducing the temperature, 
hence increasing the density to remain in pressure balance.  
This decreases the ionisation state so that bound 
transitions can exist, and line cooling dramatically 
decreases the temperature.  This leads to a sharp transition 
between a highly ionised atmosphere, heated to the Compton 
temperature, and a much more neutral disc photosphere 
\citep{kro82ionization}.  The sound speed $c_{\rm s}$ 
of the heated atmosphere gives a typical height of 
$H\sim c_{\rm s}/v_\phi(R)$, where $v_\phi$ is the 
Keplarian velocity of the disc, and produces a wind 
where $c_{\rm s}>v_{\rm esc}(R) \sim v_\phi(R)$ 
\citep{beg83wind,woo96coronae,don18wind}.  

Hence the signatures of an irradiated outer disc are 
a highly ionised X-ray atmosphere/wind, reprocessed 
optical emission, and long decays of outbursts.  
However, there is as yet no detailed
modelling of these effects and their observational tests.  
For example, the best calculations of disk-instability 
model first used the standard (unilluminated) disc models.  
These predicted that the outer regions of the disc are
self-shielded \citep{dub99irradiation}, so they remain 
unilluminated, and do not produce the observed long 
decays.  The best irradiated disk-instability models 
still use an {\it ad hoc} illumination parameter, 
such that the irradiation flux $F_{\rm Irr} = {\cal C} 
L_{\rm X}/(4\pi R^2)$ 
\citep{las01DIDNXT,dub01XNmodel}.  
The reprocessed optical emission from this can fit 
broadband optical/ultraviolet (UV) photometry of LMXBs 
\citep[e.g.,][]{gie09reprocess}, 
but gives a slightly different dependence than predicted 
by \citet{cun76irradiation}, who approximates 
the disc as isothermal and derive the well known scale 
height $H \propto R^{9/7}$ relation which leads to 
$F_{\rm Irr}\propto L_{\rm X}/R^{12/7}$ 
\citep{hyn02j1859,hyn05optuv,shi16j1655}.  

In addition, as is clear from the discussion above, the disc 
atmosphere is in no sense isothermal, and its scale height should vary 
as the illuminating spectrum changes.  This effect can be very large 
as the outburst shows a dramatic spectral transition in a way which is 
now well studied \citep{fen04uniBHBjet,rem06BHBreview}.  The spectrum 
starts off hard, dominated by Comptonisation, typically peaking at 
100~keV.  It stays hard during the rapid rise, then abruptly softens,
making a complex transition at high luminosity to a soft, disc 
dominated state, typically peaking at $2(L/L_{\rm 
Edd})^{1/4}$~keV \citep{don07XB}, before making a transition back to 
the hard state at around $L/L_{\rm Edd} \sim 0.02$ 
\citep{mac03transition}.  Since this changing spectrum gives a very 
different heating/cooling balance, the wind/atmosphere responds by 
changing its density/scale height/launch radius \citep{don18wind}.  
This predicts the strength of illumination should change during 
the outburst, especially at the spectral transition.  

We address the extent of these complexities produced 
by X-ray irradiation via modeling the changing multi-wavelength 
spectra seen in a normal black-hole binary outburst.
We use the recently developed self-consistent model (energy 
conserving) covering a wide energy range from hard-Xray to 
IR regime \citep{shi16j1655}.  
Our study enables us for the first time 
to find significant time variations in reprocessed 
fraction during outburst, to calculate how the X-ray scattered 
fraction should change, and to compare the observations and 
the predictions of reprocessing from scattering in the 
corona/wind.  
This is important for considering input values into 
the irradiated disc codes in order to predict the effect 
of X-ray illumination from a physical rather than 
phenomenological model, and for examining existing theories 
of disc instability including the effect of irradiation 
\citep{tet18irradiation}.
Thus our investigations open the way to a better understanding 
of the role of X-ray irradiation in the time-varying broadband 
spectra and light curves of LMXBs.

The amount of irradiation in the existing 
theories can only be well constrained with simultaneous 
multi-wavelength optical/UV and X-ray data.  
In this study, we use the unique coverage obtained from 
five simultaneous \textit{HST} and \textit{RXTE} spectra 
of the low galactic column black-hole binary 
XTE J1859$+$226 during its 1999--2000 outburst 
\citep{hyn02j1859}.   
Although previous works used photometric optical/UV data 
\citep[e.g.,][]{gie09reprocess}, spectroscopic data 
gives a much better determination of the continuum shape, 
which requires \textit{HST} in order to get the spectrum 
extending into the UV regime.  There are very few binaries 
for which such data exist, especially as these are often 
highly reddened due to their galactic plane location, and 
one of them is XTE J1859$+$226.  
This object is regarded as one of normal black-hole 
LMXBs in terms of the overall X-ray spectral behavior 
and light variations \cite[e.g.,][]{pei17j1859,nan18j1859}.  
Its multi-wavelength spectra spanning the brightest 
high/soft states of the outburst went through the disc 
dominated state on the decline and finally approached 
the transition to the low/hard state.  Thus all these datasets 
should be dominated by irradiated disc emission, whereas dim 
low/hard states can have additional contributions from 
cyclo-synchrotron emission from the hot electrons 
\citep{vel11sscmodel} and synchrotron emission from
the jet, which is more likely to be an infrared (IR) 
component \citep{gan11gx339jet,mer00j1118,fen04uniBHBjet}.  
These disc dominated states are the simpler
ones in which to explore the effect of reprocessing, 
and were fit to the optical/UV data by \citet{hyn02j1859}.  
However, these models were not able to simultaneously 
reproduce the X-ray spectra.  
Hence, there is no study explaining coherently the broadband 
spectral characteristics, though there are multiple observations
\citep{bro02j1859,zur02j1859,uem04j1859,cas04j1859,far13j1859,sri13j1859}.

The structure of this paper is the following.  
In Sec.~2, we display the overall X-ray and optical behaviour 
of the 1999--2000 outburst in XTE J1859$+$226, and 
in Sec.~3, we describe the data selection.  In Sec.~4, 
the details of our model are given, and the results of 
broadband spectral energy distributions (SED) analyses 
are shown in Sec.~5 and 6.  
Finally, we interpret and discuss our results 
in Sec.~7, including the results of theoretical models, 
and summarise our conclusions in Sec.~8.

\section{Overall X-ray and Optical Behavior}

\begin{figure}
\begin{center}
\includegraphics[width=8cm]{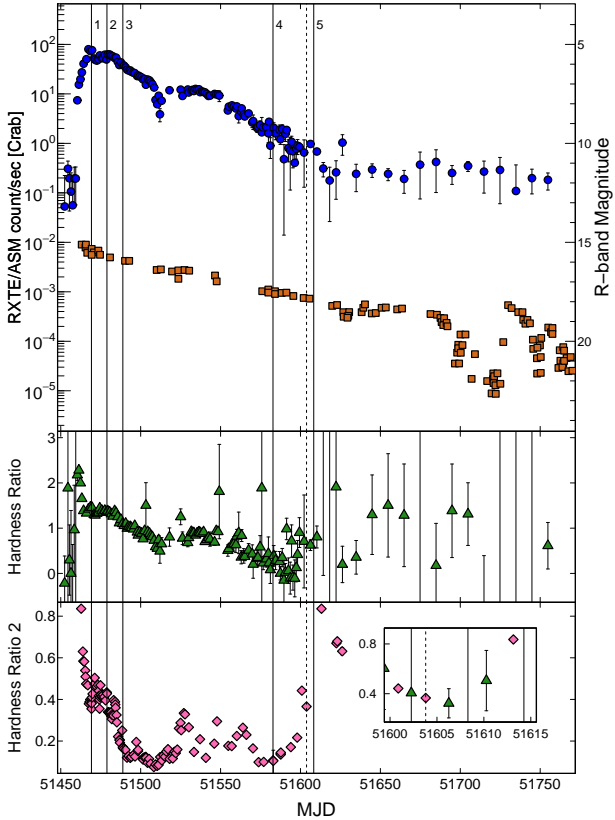}
\end{center}
\caption{Overall X-ray light curve in the 1.5--12 keV band obtained by \textit{RXTE}/ASM and overall optical $R$-band light curve digitized from Figure 2 in \citet{zur02j1859} during the 1999--2000 outburst in XTE J1859$+$226 (top panel), hardness ratio defined as 3--12~keV/1.5--3~keV in the \textit{RXTE}/ASM data (middle panel), and hardness ratio calculated defined as 6--10~keV/3--6~keV in the \textit{RXTE}/PCA data (bottom panel).  The vertical solid lines labeled by 1, 2, 3, 4, and 5 represent MJD 51469, 51478, 51488, 51582, and 51608, respectively.  The vertical dashed line shows February 29th in 2000 (MJD 51603).  The data during MJD 51600--51680 are binned every 4 days, and those after MJD 51680 are binned every 10 days.  The small window in the bottom panel shows an enlarged view of the hardness ratios during MJD 51600--51615.  Here, $-$0.3 is added to the hardness ratios of \textit{RXTE}/ASM data.}
\label{overallLC}
\end{figure}

We use the \textit{RXTE}/ASM standard products 
to look at the overall evolution of the X-ray light 
curve during the outburst (see filled circles 
in the top panel of Figure \ref{overallLC}). 
The corresponding ASM hardness ratios, defined as 
3--12~keV/1.5--3~keV (bands 2+3/band 1), are shown in
the middle panel of Figure \ref{overallLC}. 
These show that the spectrum starts in the low/hard 
state (ASM hardness ratio around 2), then makes 
a transition to a softer state during the fast rise. 
Around the maximum, the X-ray brightness shows complex 
flaring \citep[see also][]{bro02j1859}, and afterwards, 
it drops with an exponential decay with correlated 
spectral softening, interrupted by a plateau during 
MJD 51520--51545.  

The \textit{HST} data were 
taken on October 18th in 1999, October 27th in 1999, 
November 6th in 1999, February 8th in 2000, and March 5th 
in 2000, which correspond to MJD 51469, 51478, 51488, 
51582, and 51608, respectively.  
These times are indicated in Figure \ref{overallLC} 
with vertical solid lines.  Hereafter these are referred 
to as T1, T2, T3, T4, and T5.  
The UV and optical spectra of \textit{HST} were taken by 
the Space Telescope Imaging Spectrograph (STIS), and 
reduced by the standard pipeline.  The gratings used were 
G140L and G230L on the far-ultraviolet Multi-Anode 
Micro-channel Array (MAMA) detector, and G430L and G750L 
on the optical CCD.  
We put these in contexts of the optical evolution 
in the outburst by extracting the $R$-band data from 
Figure 2 in \citet{zur02j1859} (quadrangular points in
the upper panel of Figure \ref{overallLC}). These 
show a smooth decline from T1 to T5.   

The \textit{RXTE}/PCA pointed observations give 
much better spectra than the ASM, but they do not give 
continuous coverage of the outburst.  We extract 
the 3--25~keV standard product spectra, and fit these 
in XSPEC with a model consisting of 
\texttt{Tbabs*(diskbb+nthComp)} or 
\texttt{Tbabs*smedge*(diskbb+nthComp+gaussian)}, with 
hydrogen column fixed at 5.0$\times$10$^{21}$ cm$^{-2}$. 
In this paper, we use XSPEC of HEASOFT version 6.23.  
We use these models to calculate the intrinsic hardness 
ratio (unabsorbed flux in 6--10 keV/3--6 keV), shown 
in the bottom panel of Figure \ref{overallLC}.  This shows 
that there are quasi-simultaneous \textit{RXTE}/PCA data 
for T1--4, but that T5 occurs during a gap in the PCA 
data coverage.  
Importantly, this gap encompasses a state change in 
the X-ray spectrum.  The nearest PCA data taken before 
the \textit{HST} spectrum is in the soft state (PCA 
hardness ratio $\sim$0.4), while the one after is 
in the hard state (PCA hardness ratio $\sim 0.8$).  
The inset in the bottom panel of Figure \ref{overallLC} 
shows a zoom of this time period, with both the ASM 
hardness ratio (triangles, with $-$0.3 added to make them 
comparable) and the PCA hardness ratio (diamonds). 

\begin{figure}
\begin{center}
\includegraphics[width=8cm]{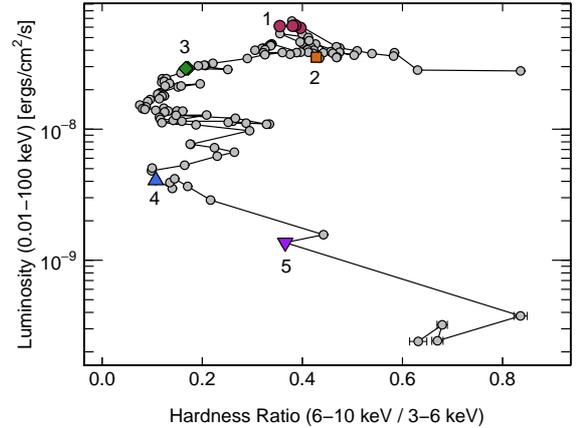}
\end{center}
\caption{Hardness-intensity diagram during the 1999--2000 outburst in XTE J1859$+$226.  The filled circles, square, diamonds, triangle, inverted triangle represent the data on MJD 51469, 51478, 51488, 51582, and 51603, respectively.  }
\label{q-diagram}
\end{figure}

Figure \ref{q-diagram} shows the outburst evolution from 
the \textit{RXTE}/PCA X-ray data on a hardness-intensity 
diagram, where we plot the unabsorbed hardness ratios 
from 6--10~keV/3--6~keV, together with the unabsorbed 
0.01--100~keV bolometric flux determined from the modelling. 
This reinforces the conclusions from the light curve and 
hardness ratios, that the source was in the very high state 
during T1--2 (red circle/orange square), then was in 
a bright soft state in T3 (green diamond), and a much 
dimmer soft state in T4 (blue triangle), while T5 (purple 
inverted triangle) was just before the transition back 
to the hard state.

\section{Selection of Multi-wavelength Spectral Data}

Our focus is to model the continuum emission over 
as broad a bandpass as possible.
We extract the \textit{HST} data from the Mikulski Archive 
for Space Telescopes (MAST) archive, 
removing any significant emission lines originating 
from the accretion disc (C IV (1550), He II (1640), 
He II (4686), H$\beta$, H$\alpha$ lines), as well as 
interstellar absorption lines (Si II, O I/Si II, C II, 
Ni III, Fe II, Fe II/Mn II, Mg II, Mg I lines), and 
the geocoronal Ly$\alpha$ line.
We bin the remaining points to obtain better 
signal-to-noise.  
%, and only used data below $\sim$8300 \AA in the optical 
%spectra to avoid issues with fringing.
We then add 4\% and 5\% systematic errors to the MAMA data 
and the CCD data of \textit{HST} spectra, respectively, 
considering the photometric accuracies.

We use the \textit{RXTE} pipeline spectral products for 
the same dates corresponding to the \textit{HST} data 
except for T5.  As for T5, we instead use the nearest 
PCA data taken 5 days before the \textit{HST} spectrum, 
considering the discussion in the preceding section.  
Also, we add 1\% systematic errors to the spectral data 
between 3--25~keV or 3--15~keV for the lower signal-to-noise 
spectra T4--5.  

We extend the wavelength coverage down to the IR regime 
by using simultaneous photometric points 
taken by the United Kingdom Infrared Telescope 
(UKIRT) 3.8-m telescope \citep{hyn02j1859}.  These 
are available for T1 ($K$ band only), while T2 and 
T5 have $J$, $H$, and $K$ bands.

\section{Multi-wavelength spectral model: {\tt optxrplir}}

Fitting the optical/UV together with the X-ray spectral 
data from an accretion flow is not straightforward. 
There is an evolving series of broadband continuum SED 
models in the literature 
\citep{gie09reprocess,sut14ULXs,shi16j1655}.  We describe 
here the most recent version 
\citep[the {\tt optxrplir} model:][]{shi16j1655} for 
completeness. Its relation to earlier models is discussed 
in \citet{shi16j1655}.

The strength of irradiation of the outer parts of 
the accretion disc by the inner flow depends on the intrinsic 
illumination pattern of the X-ray source, the shape of 
the outer disc, and its albedo. The shape of the outer disc 
is itself dependent on the X-ray illumination, with 
\citet{cun76irradiation} showing that this results in 
$H(R)\propto R^{9/7}$ so that $T_{\rm irr}(R) \propto R^{-3/7}$, 
contrasting with the standard unilluminated disc where 
$H(R) \propto R^{9/8}$ and $T_{\rm visc}(R) \propto R^{-3/4}$.  
\citet{hyn02j1859} use these relations to make a simple 
irradiated disc model, where ${T_{\rm eff}}^4(R) = 
{T_{\rm visc}}^4(R) + {T_{\rm irr}}^4(R)$.  
This is able to fit the optical/UV spectra, but does
not match the simultaneous X-ray data \citep[see Figure 5 
of][]{hyn02j1859}.  There are several reasons for this, 
firstly the model only incorporates the disc emission, although 
black-hole binaries also show a power law which extends to 
significantly higher energies than expected for an optically 
thick disc. We assume that all the emission is powered by
the standard Novikov-Thorne disc emissivity, and hence, 
the luminosity of the X-ray tail defines a radius, $R_{\rm pl}$, 
below which the gravitational power released by accretion, 
must be dissipated in optically-thin material leading to 
Comptonisation, which is characterised by an electron temperature 
$kT_{\rm pl}$ and power-law spectral index $\Gamma_{\rm pl}$, 
rather than making the highest temperature disc emission.  
The very high and intermediate states in black-hole binaries 
have additional lower temperature Comptonisation 
\citep[e.g.,][]{kub04j1550}, and hence, we similarly assume 
that this is powered by the luminosity released between 
$R_{\rm cor}$ and $R_{\rm pl}$, which is characterised by 
an optical depth, $\tau$ and electron temperature 
$kT_{\rm warm}$.  

\begin{figure*}
\begin{center}
\includegraphics[width=16cm]{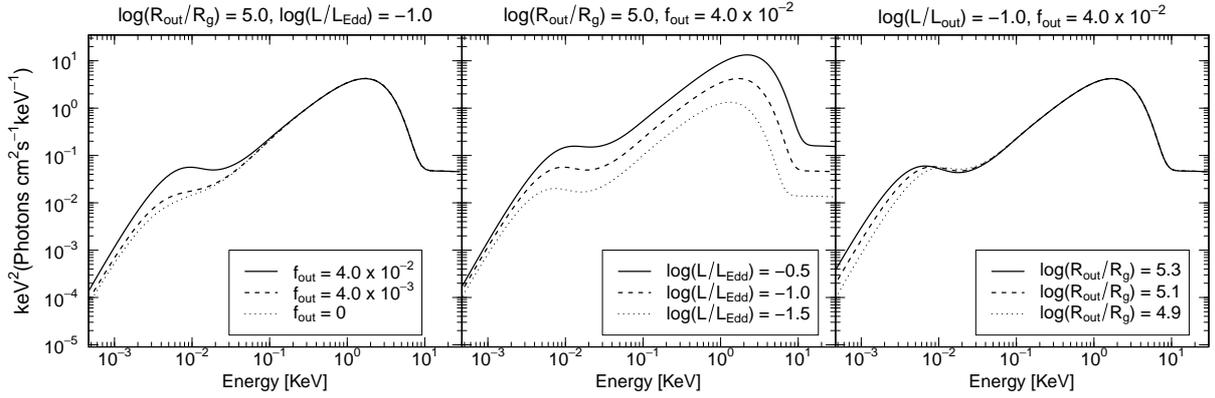}
\end{center}
\caption{Model emission of the irradiated disc in the \texttt{optxrplir} model as a function of disc height (left panel), or luminosity (middle panel), or outer disc radius (right panel).  Here, $r_{\rm cor}$ = $r_{\rm pl}$, and the parameters $r_{\rm pl}$ and $\Gamma$ are fixed to 8 and 2.0, respectively.  The other fixed parameters are the same as those described in Sec.~4.1.}
\label{testmodel}
\end{figure*}

In addition, even where the high energy X-ray emission is low, 
containing less than 10\% of the total emission, there is 
an offset in normalisation between the optical/UV emission 
from the outer disc and the X-ray emission from the inner 
disc \citep[see e.g., the upper panel of Figure 5 
in][]{hyn02j1859}.  This mismatch is due instead to 
the different temperature and density of the X-ray emitting 
inner disc compared to the optical emitting outer disc. 
The true absorption opacity can be much smaller than the
electron scattering opacity, leading to a modified rather 
than true blackbody spectrum 
\citep{sha73BHbinary,cze87accretiondisk}. 
This emission can be approximated as a colour-temperature 
corrected blackbody, 
$B_{\nu} (f_{\rm col} T_{\rm visc})/{f_{\rm col}}^4$,
where $f_{\rm col}$ is a colour temperature correction 
factor which is $\sim$1.6--2.6 at X-ray temperature 
but is $\sim$1 at optical temperature 
\citep{shi95hardening,kub01j1655,dav06spectralmodel,don12AGNs,shi16j1655}.  
We use the analytic approximation to $f_{\rm col} (T_{\rm visc})$ 
calculated in the Appendix of \citet{dav06spectralmodel} 
as implemented in \citet{don12AGNs} for each annulus with 
$R > R_{\rm cor}$.  The lower line in the left panel of 
Figure \ref{testmodel} shows the resulting spectrum without 
irradiation and with only a fairly weak, hot corona ($R_{\rm cor} 
= R_{\rm pl} = 10R_{\rm g}$) for $\log (L/L_{\rm Edd}) = -1$. 
The hot Comptonisation region has $\Gamma_{\rm pl} = 2.0$ and 
$kT_{\rm pl} = 100$~keV, while the disc is assumed to 
extend to an outer radius $R_{\rm out}=10^5R_{\rm g}$.  
The break in shape of the disc emission between the optical 
and X-ray components is clearly evident.

We incorporate irradiation by assuming about the disc height that 
$H(R) / R = f_{\rm out} (R/R_{\rm out})^{2/7}$, where $f_{\rm out}$ 
sets the fraction of luminosity from the innermost radii 
which is intercepted by the outer disc.  This is left as 
a free parameter rather than fixed to the self-consistent value 
in \citet{cun76irradiation} to allow for additional effects 
such as scattering from a wind and/or limb darkening/brightening 
of the inner disc emission. The outer disc has some albedo 
$a_{\rm out}$, so that only a fraction $(1 - a_{\rm out})$ 
of the incident flux thermalises.  We set $a_{\rm out} = 0.9$ 
on the assumption that the skin of the outer disc is
highly ionised by the X-ray illumination 
\citep{van94visualLMXB,jim02corona}. 
The left panel in Figure \ref{testmodel} shows the change 
in optical/UV emission in increasing $f_{\rm out}$ from zero 
(the intrinsic emission from the accretion flow, dotted line) 
to weak irradiation with $f_{\rm out} = 4\times10^{-3}$ 
(dashed line) to stronger irradiation with $f_{\rm out} = 
4\times10^{-2}$ (solid line).  

\textcolor{black}{In addition, irradiation also depends on luminosity 
and disc size.  
The middle panel in Figure \ref{testmodel} shows 
the expected change as a function of increasing $L/L_{\rm Edd}$ 
for fixed $f_{\rm out} = 4 \times 10^{-2}$.  
The optical reprocessed luminosity is $f_{\rm out} L$, and 
emerges as a blackbody from the fixed size outer disc, so the 
monochromatic flux in the IR/optical regime on the Rayleigh-Jeans 
tail varies only as  $\propto L^{1/4}$. This contrasts with 
the flux at UV and higher energies which changes by 
a much larger factor.  
The right panel of Figure \ref{testmodel} shows instead the effect of
changing the size scale of the outer disc for fixed luminosity and
reprocessed fraction. This changes the Rayleigh-Jeans IR/optical
tail, with almost no effect on the emission at higher energies.
Thus the UV flux is most sensitive to the reprocessed fraction, the
UV and X-ray emission together is most sensitive to the intrinsic 
luminosity, while the IR/optical emission is most sensitive to 
the size of the outer disc. }

\section{Constraining Parameters in Broadband fitting with {\tt optxrplir}}

The model \texttt{optxrplir} contains the black-hole mass 
and the distance to the object as free parameters.  
We constrain the range of the black-hole mass to 
6.16--15.5 $M_{\odot}$ \citep{cor11j1859}
and fix the distance of 8~kpc 
\citep{hyn05optuv,tet16BHXBdata}.  
The black-hole spin parameter $a_{*}$ is fixed to 0.  
This model assumes a 60-deg inclination in computing 
the disc spectrum.  We fix the normalization to 1, 
since the inclination of this object is suggested to be 
close to 60 deg \citep{cor11j1859}. At this inclination 
and spin, the Doppler boosting of the inner disc emission 
is almost completely offset by gravitational redshift, 
and hence, we do not impose any relativistic corrections. 
We also fix the temperature of the hot corona at 100 keV, 
which is much higher than the energy range of 
\textit{RXTE}/PCA data.

%There can also be reflection of the Comptonisation 
%components from the inner disc.  
\textcolor{black}{The hard X-rays from the corona can directly 
illuminate the inner disc as well as the outer disc. 
Some fraction of these are scattered (reflected) by electrons in the
disc, while the remainder is absorbed by bound-free transitions
which can be followed by fluorescent line emission. They combine and 
produce a reflected spectrum which rises sharply from 1-10~keV as the
absorption opacity decreases, but with a strong iron edge and emission
line superimposed. This can be approximately modelled 
using a smeared edge (\texttt{smedge}), with the 
cross-section index fixed at $-$2.67. We also include a gaussian line
where required by the data.
We model the companion star as a constant blackbody component 
(\texttt{bbodyrad}). This is a late-type K star according to 
\citet{cor11j1859}, and with temperature 
of 3.2--3.6$\times$10$^{-4}$ keV and radius of 
4.5--6.2$\times$10$^{5}$ km \citep{all73quantities}.}

To fit the raw data requires that we include reddening 
and absorption from dust and gas in the interstellar 
medium (ISM). We use the \texttt{phabs} model for the gas 
phase as it does not include the neutral edges below 
13.6~eV (e.g., from Fe I).  These are not present in our 
data at the expected level from neutral material due to 
the multiphase nature of the ISM, where most Fe is ionised 
to Fe II.  We constrain 
$N_{\rm H}$ in the gas phase to the range 
1.0--3.0$\times$10$^{21}$ cm$^{-2}$ by using \textit{ASCA}/GIS 
data on MJD 51474. This extends down below 1~keV and is 
more sensitive to absorption than the \textit{RXTE}/PCA data.

\textcolor{black}{
There is a range of UV reddening curves even for a fixed 
$R_V = 3.1$ parameter which is the ratio of total to selective
extinction at the optical $V$ band 
\citep{fit99extinction,car89extinction,sea79extinction}.  
These differ at the 20\% level around the 2200\AA\ feature 
and at the shortest wavelengths (less than 1500\AA) (see e.g., 
Figure 1 of \citealt{fit99extinction}, or Figure 9 of 
\citealt{fit07extinction}).}
\citet{hyn05optuv} used the reddening curve of 
\citet{fit99extinction}, but this is not available in 
XSPEC. Instead, we use the \citet{sea79extinction} 
reddening model (\texttt{uvred} in XSPEC) as being the closest 
to the \citet{fit99extinction} curve around the 2200\AA\ 
feature. 
This only extends down to 3700\AA (3.4~eV), 
so we first fit all five datasets simultaneously, 
using a restricted energy range of 0.0034--25~keV with 
the model \texttt{uvred*phabs*smedge*(optxrplir)}, tying 
the extinction, column density and black hole mass across 
all datasets.  
We find best-fit values for these of 
$E(B-V) = 0.526$, $N_H = 3.0\times 10^{21}$~cm$^{-2}$
and mass of 6.9 $M_{\odot}$. 
\textcolor{black}{
The best-fit value of reddening is slightly different 
to the value of $E(B-V)=0.58\pm 0.12$ derived by 
\citet{hyn02j1859}, but this is due to the difference 
in shape of the reddening curve \citet{fit99extinction}.  
The effect of our reddening correction with $E(B-V)=0.526$ 
from \citet{sea79extinction} is very similar to that of 
$E(B-V)=0.58$ from \citet{fit99extinction}.
We use the extinction law from \citet{car89extinction} 
to extend the reddening down to the optical, having 
checked that the difference is within a few percents between 
\citet{sea79extinction} and \citet{car89extinction}
in the region above 3700\AA\ \citep{fit07extinction}.  }

\textcolor{black}{
We then deredden the optical/UV/IR spectra with the best-fit 
value of $E(B-V)$, and fit all 5 spectra simultaneously 
across the entire band with the model 
\texttt{phabs*smedge*(bbodyrad+optxrplir)}, 
to obtain the best-fit values for the parameters of 
the companion star.  The best-fit temperature and 
radius are estimated to be 3.6$\times$10$^{-4}$ keV 
and 6.2$\times$10$^{5}$ km, respectively.  
}

We note that the black hole mass, as well as being 
consistent with previous determinations, gives the luminosity 
just before the soft-to-hard transition of around 1--2\% of 
the Eddington luminosity, as expected from 
the advection-dominated accretion flow models of the low/hard 
state \citep[ADAF;][]{nar95ADAF}, and in line with what is 
generally observed \citep{mac03transition}.

\section{Evolution of the Accretion Flow}

In this section, we fit the spectra at each epoch 
(T1--5) individually, using the deredden optical/UV data 
together with the\textit{RXTE}/PCA X-ray data.  
We again fit with the model 
\texttt{phabs*smedge*(bbodyrad+optxrplir)}, but now have 
all the system parameters at their best-fit 
values as determined in the preceding section in order to better 
constrain the remaining parameters of the accretion flow.  
%The {\tt phabs} absorption applies only to the 
%X-ray data to correct this X-ray absorption.  
We give the de-convolved, de-absorbed SED in the top 
window in Figure \ref{SED1}, and show residuals 
to the best fit model in the small window below each 
spectrum.  
The best-fit parameters and their errors are given 
in Table \ref{parameter}.

\subsection{Modelling of the Very High State at the Early Stage}

The hardness-intensity diagram (see Figure \ref{q-diagram})
clearly shows that the X-ray spectra in T1 and T2 are in 
the very high state. This state can show strong 
low-temperature and optically-thick thermal Comptonisation 
as well as a power-law tail to higher energies indicating 
a hotter, optically-thin component 
\citep[e.g.,][]{zyc01diskcorona,kub01j1655,kub04j1550}. 
We deal with this component by allowing $R_{\rm cor}$ 
to be a separate free parameter, along with the electron 
temperature and optical depth of this component.  

\begin{figure*}
\begin{center}
\begin{minipage}{.49\textwidth}
\includegraphics[width=7.5cm]{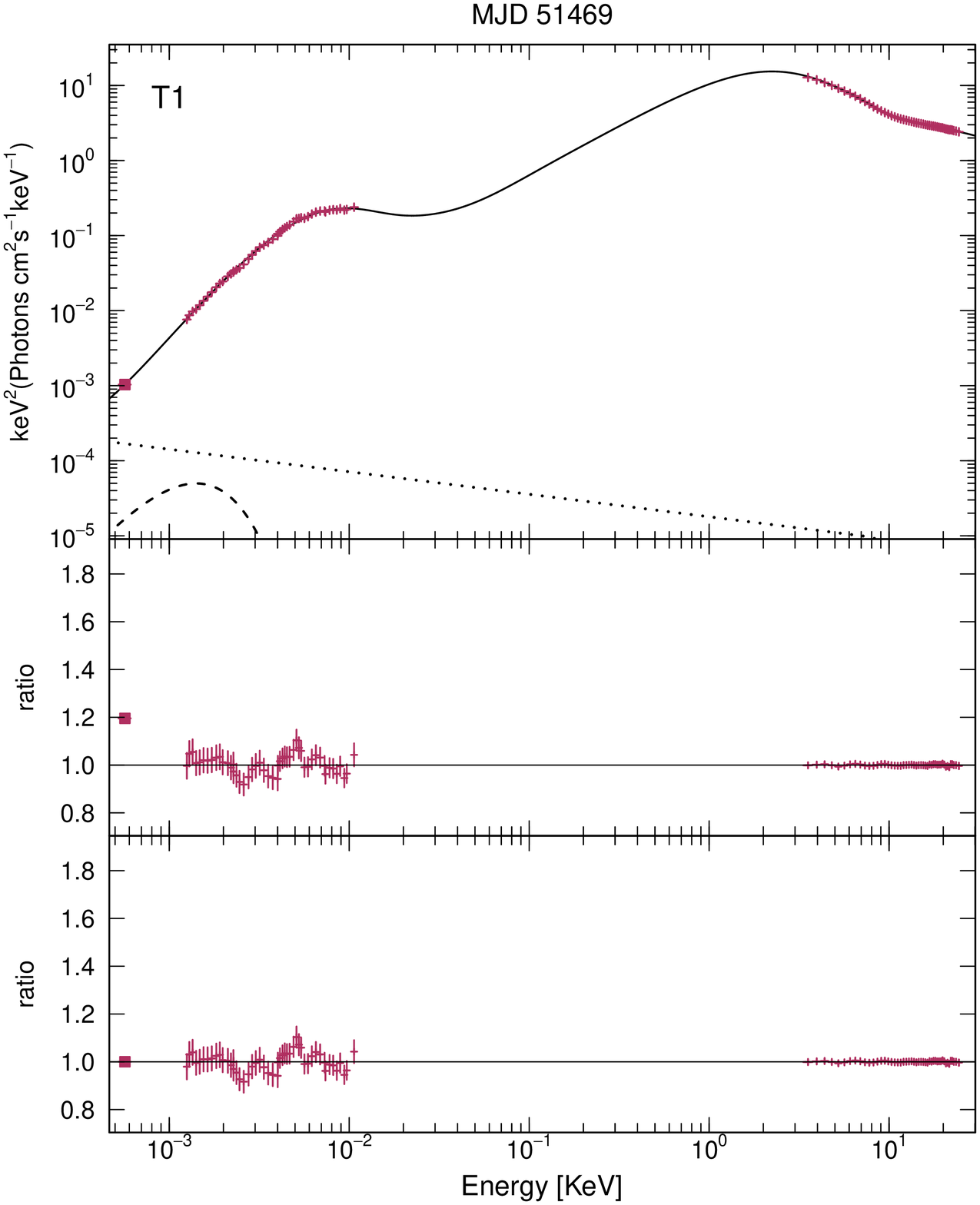}
\end{minipage}
\begin{minipage}{.49\textwidth}
\includegraphics[width=7.5cm]{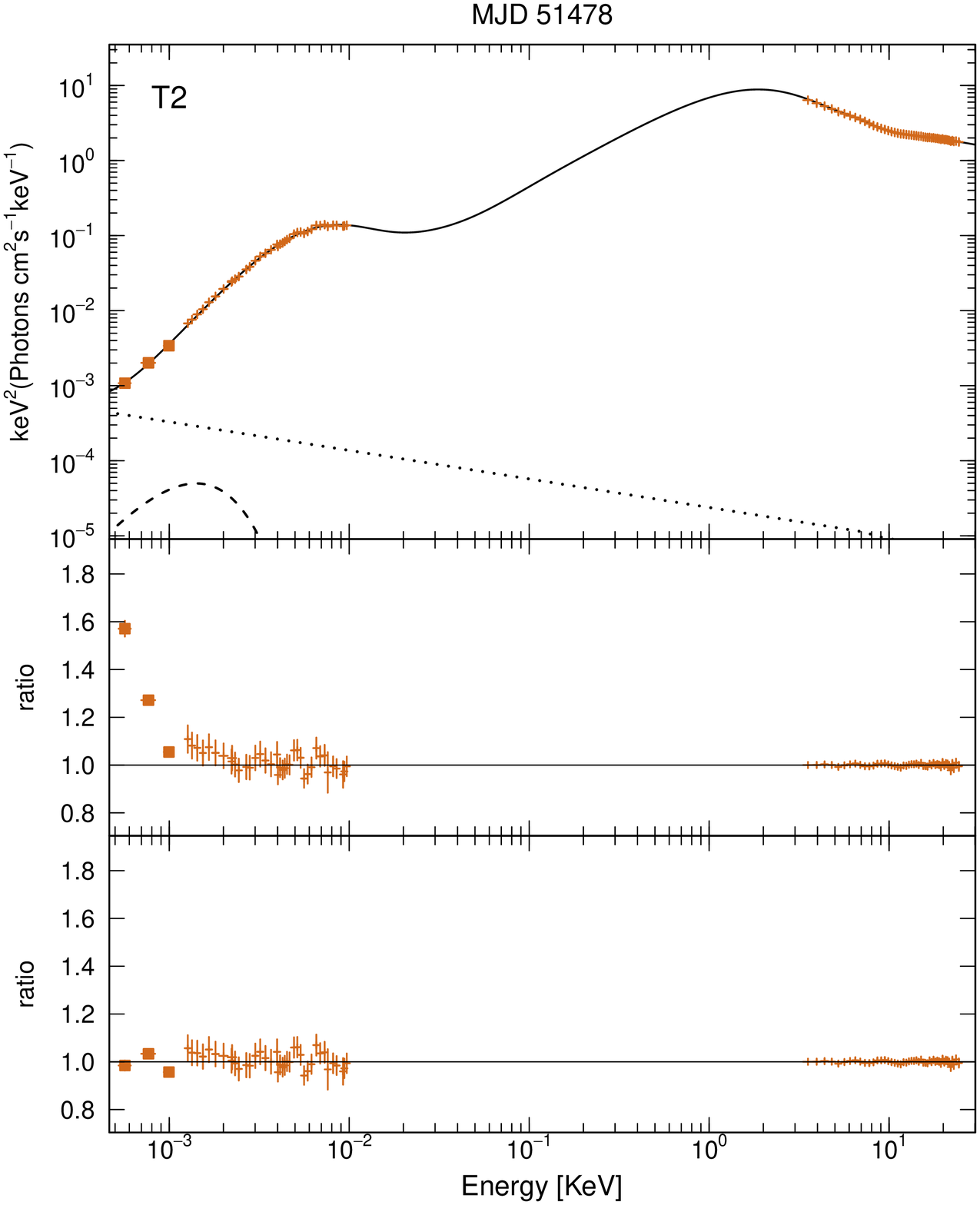}
\end{minipage}
\\
\begin{minipage}{.49\textwidth}
\includegraphics[width=7.5cm]{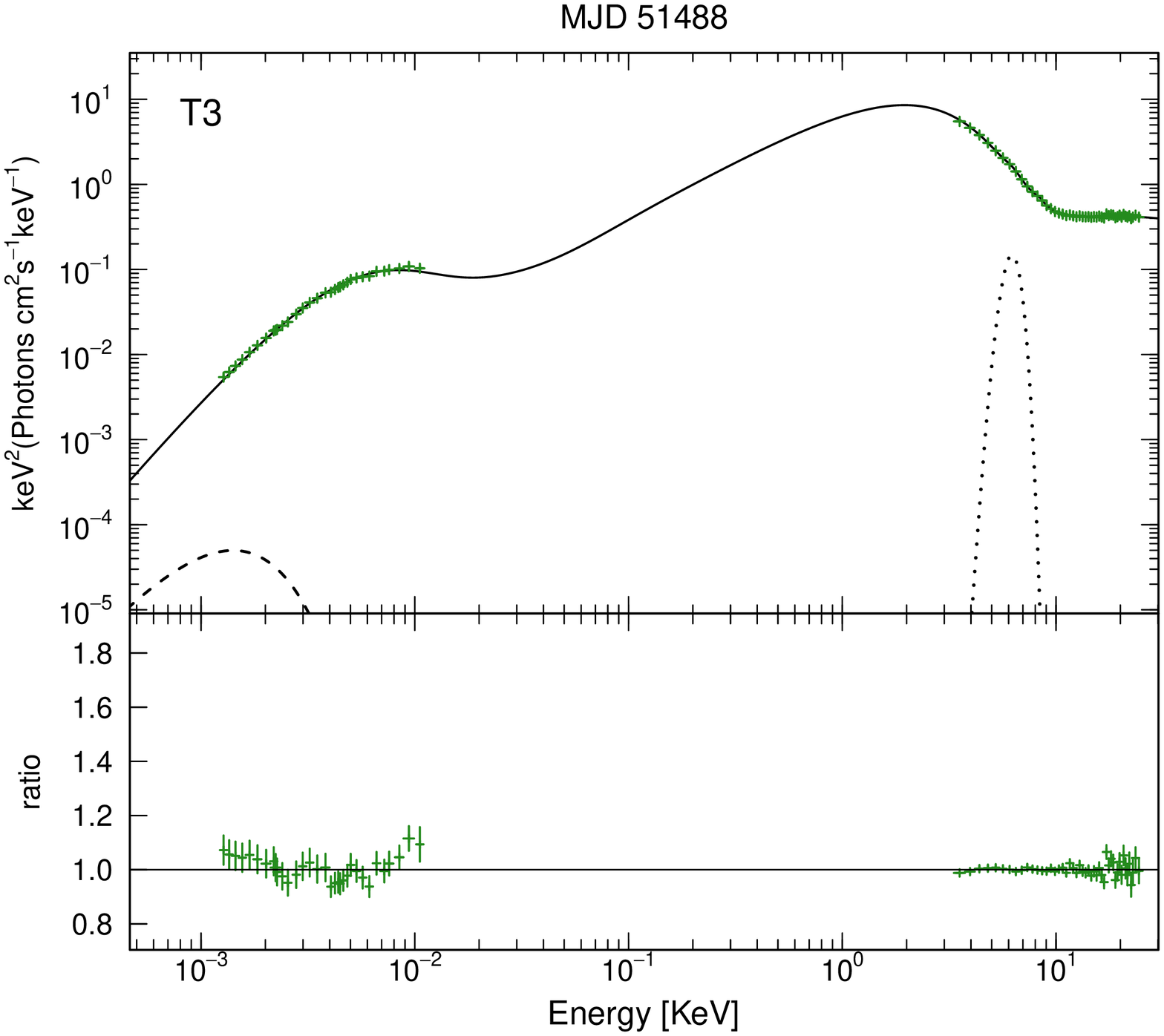}
\end{minipage}
\begin{minipage}{.49\textwidth}
\includegraphics[width=7.5cm]{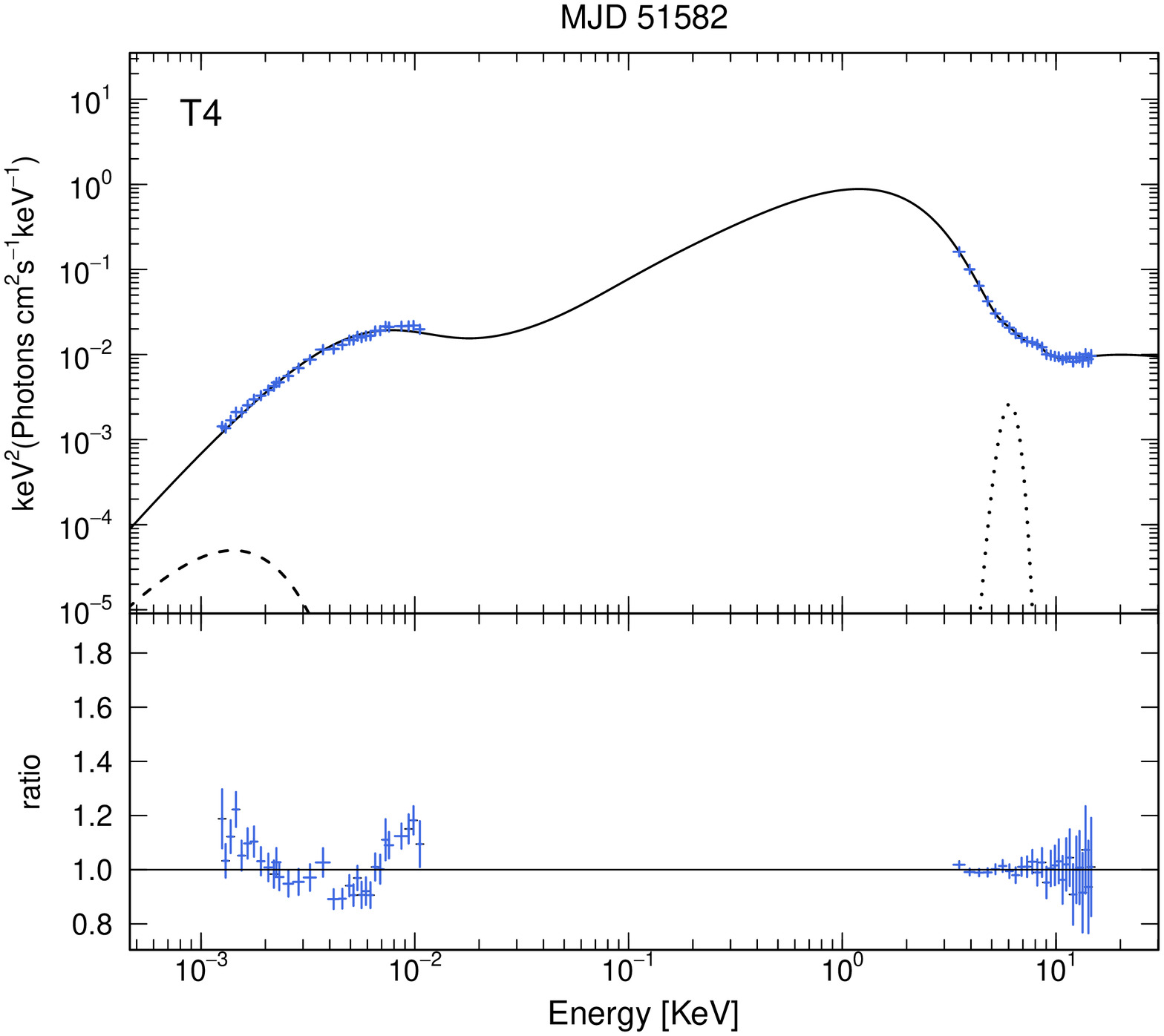}
\end{minipage}
\\
\includegraphics[width=7.5cm]{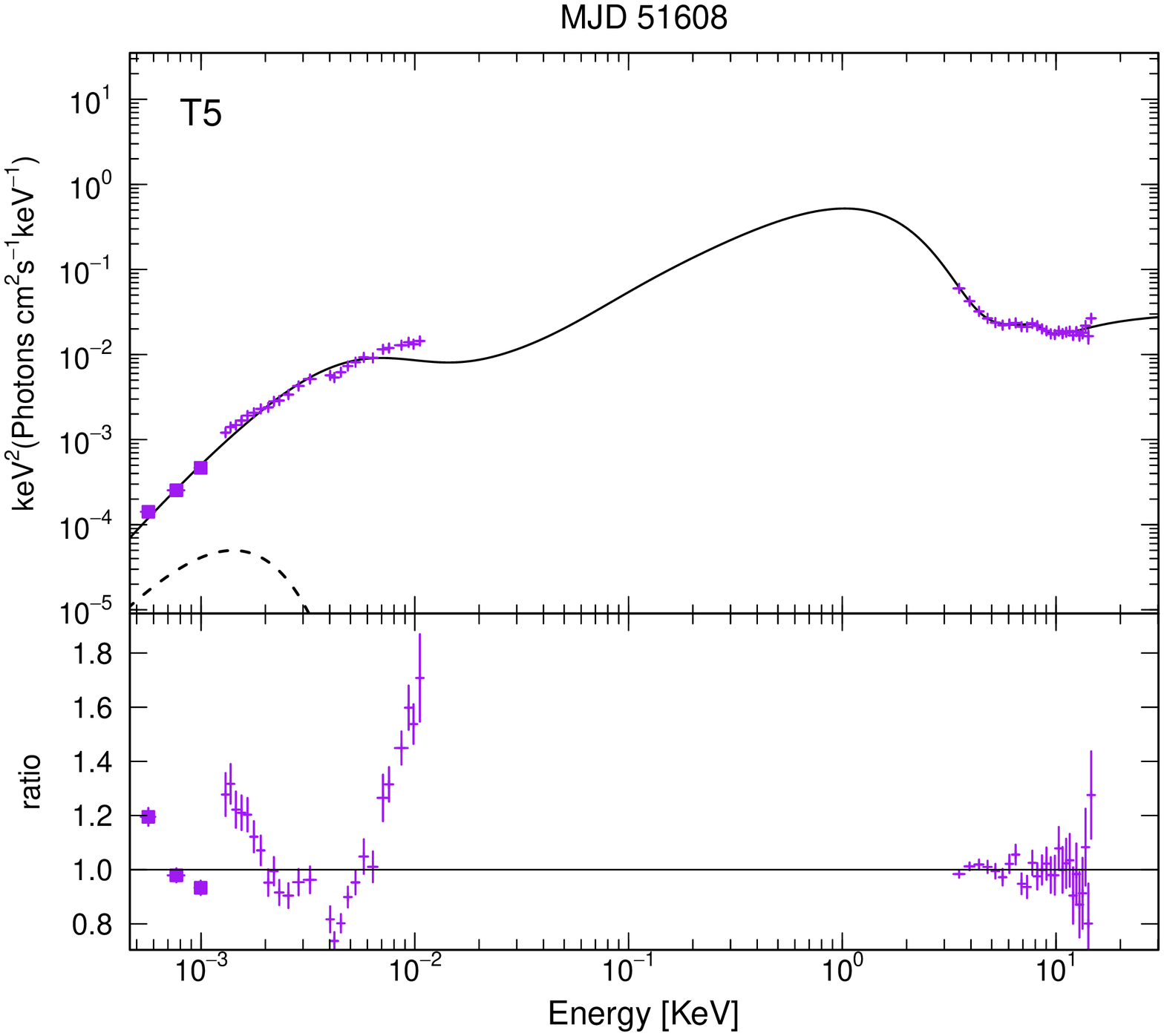}
\end{center}
\caption{Multi-wavelength SEDs and their ratios on T1--T5.  On T1 and T2, we prepare the ratio without a power-law model for comparison in the middle windows.  The crosses and solid line represent the observed SED and the emission reproduced by our model, respectively.  The rectangles are the IR data.  The dashed line represents the emission from the companion star.  The dot line in the upper panels represents the model emission from synchrotron jet ejection, and that in the middle panels represents the model emission from iron line.  }
\label{SED1}
\end{figure*}

The initial fit of this baseline model to T1 and T2 shows 
an excellent match to the optical/X-ray spectra, but 
with a strong IR excess.  
\textcolor{black}{This is most clearly seen in the residuals 
as for T2, which are displayed in the middle window of 
the upper right panel of Figure \ref{SED1},} where there are 
IR data across $J$, $H$ and $K$ bands. This could indicate 
a contribution from the radio jet. This is known to 
closely follow the hard X-ray Comptonised flux, even into 
the soft states \citep{zdz11cygX1}. The very high and 
intermediate states are often the ones with the strongest 
jet emission \citep{fen04uniBHBjet}, and radio monitoring of 
XTE J1859$+$226 during this outburst showed that the radio 
luminosity was largest in these initial stages of the outburst 
\citep{bro02j1859}.  
We thus add a power law to approximately model 
the synchrotron jet emission. This gives a much better fit 
to the IR region, without changing the fit at higher energies 
(see the lower windows of the upper two panels in Figure 
\ref{SED1}).  
\textcolor{black}{By comparing the middle and bottom panels 
in the upper two panels of Figure \ref{SED1}, we can see 
a significant jet contribution to the IR flux, with 
a rather steep synchrotron emission index of $\sim$2.5, 
which is consistent with the IR jet emission seen 
in GX 339$-$4, another black-hole LMXB \citep{gan11gx339jet}.  
As for the IR excess in T2, the contribution of jet ejection 
is about 20\%, which is estimated from the model flux of 
the power-law component on XSPEC.} 

\subsection{Modelling of the High/Soft States}

The X-ray spectra in T3 and T4 are clearly in the high/soft 
state according to the hardness-intensity diagram 
(see Figure \ref{q-diagram}).  Hence we do not include 
the low-temperature Comptonization component that is 
considered in modelling of SEDs in the very high state 
in the previous subsection, i.e., we set $R_{\rm pl} = 
R_{\rm cor}$. 
We also now include a gaussian line with upper limit 
on the width of $0.5$~keV as there are small but significant 
residuals in the 6--7~keV bandpass which indicate that 
the phenomenological \texttt{smedge} model is not sufficient 
to describe the reflection spectrum from the inner disc 
in these data. 

The best-fit SEDs for T3 and T4 are shown in the middle 
panels in Figure \ref{SED1}.   
The X-ray spectra are well reproduced by dominant disc 
emission, with a weak power-law Comptonised tail, and 
a small contribution from the iron line for both high/soft 
state SEDs. 
The UV and optical spectra for T3, taken only 10 days 
after the very high state and at similar luminosity, are 
well fit by the expected irradiated disc emission.  
However, the optical/UV spectra of T4, taken about three months 
after T3 with order of magnitude lower luminosity, show 
small but significant deviations away from the irradiated 
disc model, despite both the outer disc radius and height 
being free to vary. The best-fit values of both these 
parameters are different from those of T1--3, with the disc 
outer radius being significantly smaller and the disc height 
being significantly larger. The fit is, however, quite
poor, with $\chi^2/$~dof larger than 2, and thus, the error 
ranges on the parameters may not be reliable.

\subsection{Modelling near the Transition to the Low/Hard State}

The spectrum T5 is the lowest luminosity spectrum, taken 
just before the transition back to the low/hard state 
(Figure \ref{q-diagram}). 
This is adequately fit using only the {\tt smedge} 
to model the reflection features, and does not require 
the separate Fe line emission. 
The resulting SED is represented in the bottom panel of 
Figure \ref{SED1}, and we can see the multi-wavelength fit 
is very poor.  

Given that the X-rays are not strictly simultaneous 
in these data, and that the source was clearly in the low/hard 
state 4 days later, it is possible that the source was 
already in the low/hard state during the time of the T5 spectra. 
Thus there could be a potential component from the jet 
as well as from an irradiated disc. However, the jet should 
contribute more in the IR emission, similar to T1--2, whereas 
the IR flux level is clearly fairly well matched by the data.  

\textcolor{black}{
Instead, it is the UV flux which is most discrepant, being 
underpredicted by almost a factor 2 at the shortest wavelengths. 
This rules out an origin in either the jet or from uncertainties 
in the K-type companion star luminosity or temperature 
as both of these are red components.
This is because the $R$-band magnitude of the companion star 
must be dimmer than the minimum of $R=23$ seen in the quiescent 
intervals in Figure \ref{overallLC}, which is 100 times fainter 
than the $R=18$ seen during T5.
Also, there is a small $\sim 20$\% discrepancy between the IR and 
reddest \textit{HST} optical points but this could be due to 
time variability as these data are not strictly simultaneous.
The small discontinuity between the red and blue 
\textit{HST} spectra may be also due to variability. }

\textcolor{black}{
The derived best-fit values of disc outer radius and height are
different from those of T1--3, with the disc outer radius being
significantly smaller, similar to T4.  However, the irradiated 
disc models now do not match well to the observed UV 
spectral curvature.  The shape of the UV spectrum has hardened 
considerably (see next section, and see the discussion of 
\citealt{hyn02j1859}) and is now bluer than predicted by 
the irradiated disc models. 
Since this is a change in curvature, it cannot be produced 
by uncertainties in the absolute reddening curve, but instead 
is showing there is an intrinsic change in the shape of 
the UV spectrum. }

\ifnum0=1
\textcolor{black}{
Although the contribution of the companion star is about 10\% 
at IR wavelengths, the irradiated disc component is still 
dominant.  Only the companion star cannot explain the high 
optical/UV flux because it is a K-type star \citep{cor11j1859}.  }

Again, the best-fit values of disc outer radius and height are
different from those of T1--3, with the disc outer radius being
significantly smaller, similar to T4.  However, as shown by the
residuals, the fit is now extremely poor, with $\chi^2/$~dof $\sim
10$. Some of this may be due to intrinsic variability giving 
a different normalisation between the four \textit{HST} wavelength 
ranges, but plainly the main issue is the lack of spectral 
curvature in the optical/UV regime compared to that predicted 
for an irradiated disc. This change in curvature in 
the optical/UV spectra for T1--3 compared to T5 is irrespective 
of the X-ray data, e.g., see the de-reddened \textit{HST} 
spectra of Figure 4 in \citet{hyn02j1859}.
\fi

\begin{table*}
	\centering
	\caption{Best-fit parameters of our modelling in the 5 sets of broad-band spectra.  }
	\label{parameter}
	\begin{tabular}{ccccccc}
		\hline
		Model & Parameter & T1 & T2 & T3 & T4 & T5\\
		\hline
		\hline
smedge & $E_{\rm c}$$^{*}$ & 7.6$\pm$1.1 & 7.4$\pm$1.0 & 8.4$\pm$0.3 & 8.3$\pm$1.7 & 8.1$\pm$0.6 \\
       & $f$$^{\dagger}$ & 0.29$\pm$0.71 & 0.5$\pm$0.5 & 1.0$\pm$0.3 & 1.0$\pm$0.7 & 1.0$\pm$0.6 \\
       & $W$$^{\ddagger}$ & 4.7$\pm$34 & 5.6$\pm$11.3 & 4.5$\pm$1.4 & 2.3$\pm$1.2 & 2.7$\pm$2.2 \\
\hline
gaussian & $E_{\rm l}$$^{\S}$ & -- & -- & 6.1$\pm$0.3 & 6.0$\pm$0.4 & -- \\
         & $\sigma$$^{\P}$ & -- & -- & 0.50$\pm$0.06 & 0.5$\pm$0.5 & -- \\
         & N$_1$$^{|}$ & -- & -- & (4.9$\pm$1.5)$\times$10$^{-3}$ & (0.9$\pm$6.0)$\times$10$^{-5}$ & -- \\
\hline
pegpwrlw & $\alpha$$^{**}$ & 2.3$\pm$2.8 & 2.4$\pm$0.6 & -- & -- & -- \\
         & N$_2$$^{\dagger\dagger}$ & 0.65$\pm$0.53 & 1.4$\pm$0.7 & -- & -- & -- \\
\hline
optxrplir & log($L$/$L_{\rm Edd}$)$^{\ddagger\ddagger}$ & $-$0.38$\pm$0.09 & $-$0.59$\pm$0.02 & $-$0.68$\pm$0.01 & $-$1.67$\pm$0.01 & $-$1.88$\pm$0.02 \\
          & $R_{\rm cor}$$^{\S\S}$ & 32$\pm$50 & 22$\pm$300 & -- & -- & -- \\
          & log($R_{\rm out}$/$R_g$)$^{\P\P}$ & 5.26$\pm$0.01 & 5.23$\pm$0.02 & 5.20$\pm$0.01 & 4.92$\pm$0.02 & 4.91$\pm$0.01 \\
          & $kT_{\rm es}$$^{\parallel}$ & 1.06$\pm$0.16 & 1.1$\pm$0.4 & -- & -- & -- \\
          & $\tau$$^{***}$ & $\gtrsim$7.3 & $\gtrsim$4.7 & -- & -- & -- \\
          & $R_{\rm pl}$$^{\dagger\dagger\dagger}$ & 19$\pm$9 & 19$\pm$1 & 10.30$\pm$0.06 & 8.22$\pm$0.25 & 10.81$\pm$0.57 \\
          & $\Gamma$$^{\ddagger\ddagger\ddagger}$ & 2.42$\pm$0.07 & 2.30$\pm$0.06 & 2.17$\pm$0.04 & 2.16$\pm$0.24 & 1.83$\pm$0.13 \\
          & $f_{\rm out}$$^{\S\S\S}$ & (4.4$\pm$0.8)$\times$10$^{-2}$ & (4.0$\pm$0.3)$\times$10$^{-2}$ & (3.2$\pm$0.1)$\times$10$^{-2}$ & (6.2$\pm$0.3)$\times$10$^{-2}$ & (4.0$\pm$0.4)$\times$10$^{-2}$ \\
\hline
\hline
$\chi^2/$~dof & & 0.47 & 0.63 & 0.99 & 2.26 & 10.47 \\
\hline
\multicolumn{7}{l}{$^{*}$Threshold energy of reflection in units of keV.}\\
\multicolumn{7}{l}{$^{\dagger}$Maximum absorption factor at the threshold energy.}\\
\multicolumn{7}{l}{$^{\ddagger}$Smearing width in units of keV.}\\
\multicolumn{7}{l}{$^{\S}$Line energy in unit of keV.}\\
\multicolumn{7}{l}{$^{\P}$Line width in unit of keV.}\\
\multicolumn{7}{l}{$^{|}$Total photons/cm$^{-2}$/s in the line of sight.}\\
\multicolumn{7}{l}{$^{**}$Photon index of power law $\alpha$.}\\
\multicolumn{7}{l}{$^{\dagger\dagger}$Flux in units of $10^{-12}$ ergs/cm$^2$/s over the energy range (0.0005--0.02 keV).}\\
\multicolumn{7}{l}{$^{\ddagger\ddagger}$Luminosity divided by the Eddington luminosity in logarithmic scales.}\\
\multicolumn{7}{l}{$^{\S\S}$Radius of the low-temperature Comptonization component in units of $R_g$.}\\
\multicolumn{7}{l}{$^{\P\P}$Disc radius in units of $R_g$ in logarithmic scales.}\\
\multicolumn{7}{l}{$^{\parallel}$Temperature of the low-temperature Comptonization component in units of keV.}\\
\multicolumn{7}{l}{$^{***}$Optical depth in the low-temperature Comptonization component.}\\
\multicolumn{7}{l}{$^{\dagger\dagger\dagger}$Radius of the central hot corona in units of $R_g$.}\\
\multicolumn{7}{l}{$^{\ddagger\ddagger\ddagger}$Photon index of the central hot corona.}\\
\multicolumn{7}{l}{$^{\S\S\S}$Geometry-dependent factor of the thermalized fraction at the illuminated outer disc.}\\
\end{tabular}
\end{table*}

\subsection{Changing Optical/UV Spectral Shape}

\begin{figure}
\begin{center}
\includegraphics[width=8.0cm]{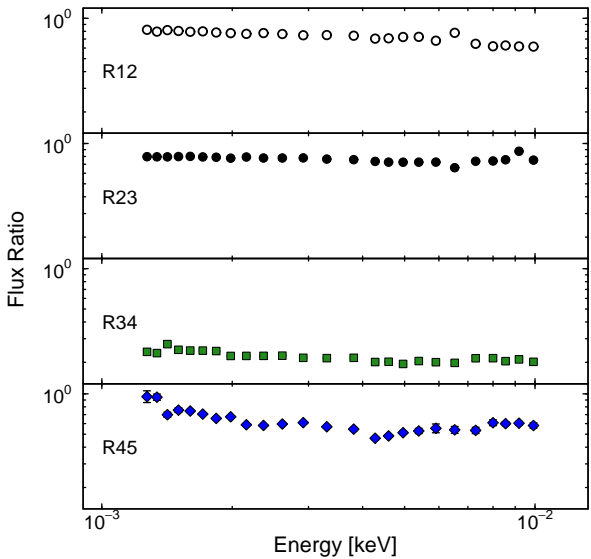}
\end{center}
\caption{Flux ratio of the UV and optical spectra taken by \textit{HST} between two adjacent time zones.  
%For visibility, we add $-$0.15, 0.15, and $-$0.52 to R2, R3 and R4, respectively.  
}
\label{fluxratio}
\end{figure}

We focus on the changing optical/UV spectral shape 
between the classic irradiated disc shape in T1--3 
\citep[described as UV soft spectra in][]{hyn02j1859} 
to the less curved shape of T4 and especially T5 
\citep[UV hard,][]{hyn02j1859}.  We look at this in 
a model independent way by computing the ratio of 
optical/UV data from the accretion flow in each
epoch.  We subtract the contribution of the companion 
star from each spectrum, and then estimate the ratios 
between the resultant 
fluxes.  The ratio of T2/T1 is R12, the ratio of T3/T2 
is R23 etc.  These flux ratios are shown as a function 
of energy in Figure \ref{fluxratio}.  These show that 
R12 and R23 are slightly decreasing with increasing 
energy across the optical/UV band as expected for 
constant outer disc radius with slightly decreasing 
illuminating flux and bolometric luminosity (see
also the left and middle panels of Figure \ref{testmodel}).

The large drop in flux between T3 and T4 is evident 
in the much lower normalisation of R34, but the flux 
ratio between these two widely separated epochs is 
unexpectedly fairly constant with energy.  
The model spectra shown in the middle panel of 
Figure \ref{testmodel} show how the irradiated disc 
spectrum in the \texttt{optxrplir} model changes with 
changing only log($L/L_{\rm Edd}$), and this clearly 
makes much more impact on the UV flux than on the optical 
flux.  
%; however, the R34 ratio is mostly flat across this 
%entire range. 
Instead, in the context of the irradiated disc models, 
to decrease the optical flux requires a smaller 
$R_{\rm out}$ (see also the right panel of Figure 
\ref{testmodel}). 

\begin{figure*}
\begin{center}
\begin{minipage}{.49\textwidth}
\includegraphics[width=8.0cm]{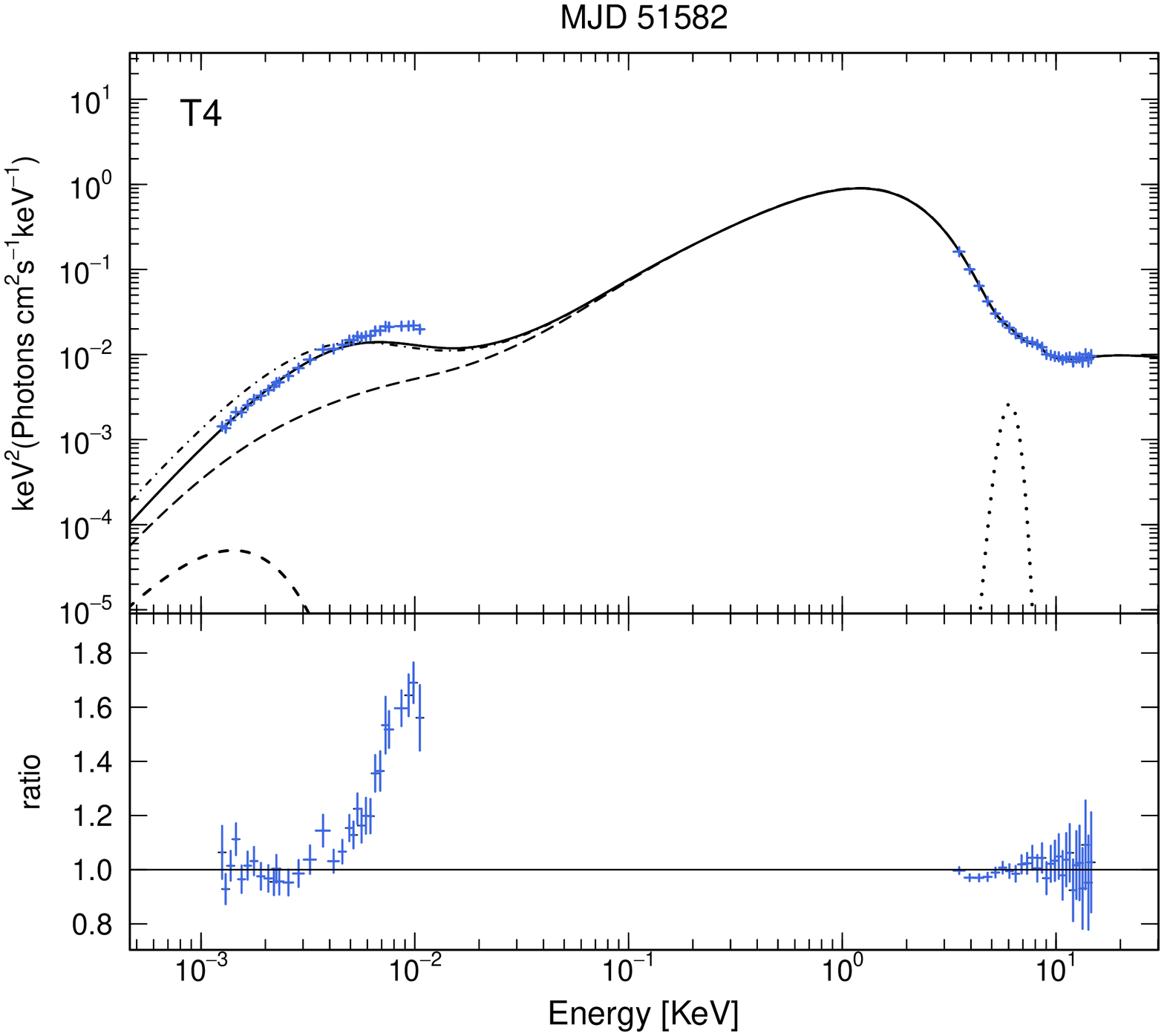}
\end{minipage}
\begin{minipage}{.49\textwidth}
\includegraphics[width=8.0cm]{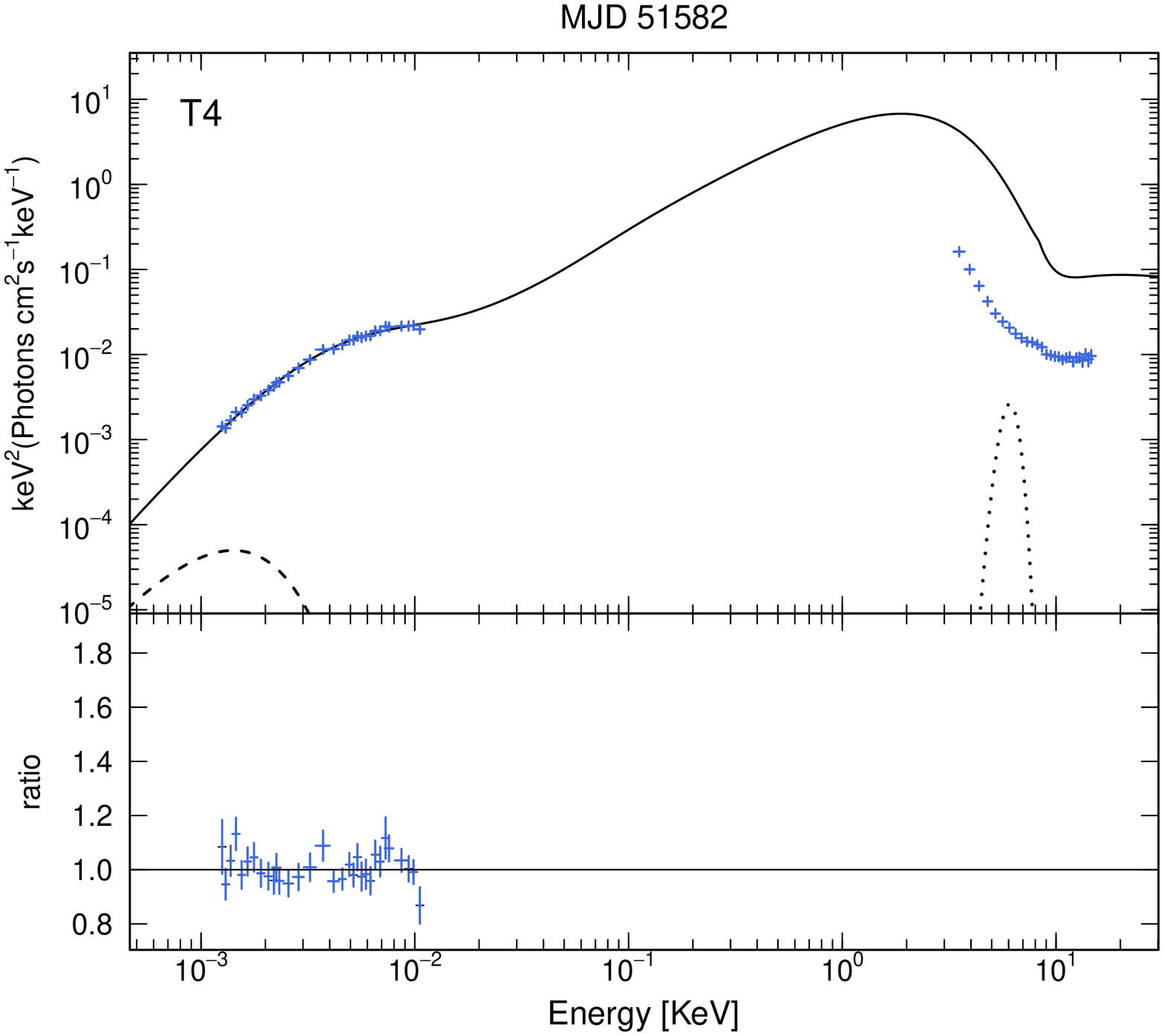}
\end{minipage}
\end{center}
\caption{Multi-wavelength SED modeling and the ratio of SED on T4 with $f_{\rm out}$ fixed to 4.0$\times$10$^{-2}$ (left panel), and those without the X-ray spectra (right panel).  The crosses and solid line represent the observed SED and the emission reproduced by our model, respectively.  The dashed and dot lines represent the emission from the companion star and that from iron line, respectively.  The dash-dotted line in the left panel is the model with log($R_{\rm out}$/$R_{\rm g}$) = 5.2.  The long-dashed line in the left panel represents an unirradiated disc.  }
\label{SED4-imp}
\end{figure*}

We explore whether a decrease in log($L/L_{\rm Edd}$) and 
$R_{\rm out}$ can explain most of the observed change 
in optical/UV emission in T4.  We fit the SED in T4 with 
$f_{\rm out}$ fixed to 4.0$\times$10$^{-2}$ similar to 
T1--3.  While the resulting fit, shown in the left panel 
of Figure \ref{SED4-imp}, has the same drop in 
log($L/L_{\rm Edd}$) and log($R_{\rm out}/R_{\rm g}$) 
as in the fit with a free $f_{\rm out}$, this gives 
a better match to the optical data than the fit allowing 
$f_{\rm out}$ to be free (the right middle panel of 
Figure \ref{SED1}), but a dramatically worse fit to 
the UV spectra (see the left panel of Figure~\ref{SED4-imp}). 
This mismatch is now similar to that of T5 in the previous 
section.  
The dashed-dotted line shows a comparison model where 
log($R_{\rm out}$) is fixed at 5.2 as in T3, showing how 
the optical data strongly require a decrease in $R_{\rm out}$ 
to compensate for the change in log($L/L_{\rm Edd}$).  
This strongly overpredicts the observed optical data. 

\ifnum0=1
The \textit{HST} data in T4 can instead be fit by 
ignoring the simultaneous X-ray data, i.e., by assuming 
that the disc is not in steady state.
The right panel of Figure~\ref{SED4-imp} shows 
how the curvature across the entire optical/UV bandpass 
can be fit by assuming very weak illumination of 
a disc with a higher mass accretion rate than that 
consistent with the X-ray emission.  The curvature 
in the UV region is then almost entirely from the increasing 
colour-temperature correction rather than from irradiation.  
The disc outer radius is again significantly smaller than 
in T1--3.  The estimated parameters in these additional 
two SED fittings are given in Table \ref{parameter2}.  
Thus there is a potential solution to explain the shape of
the \textit{HST} spectrum in T4 if the outer disc is 
intrinsically hotter than that expected from the mass 
accretion rate as measured from the inner disc. 
\fi

All of these models assumed that the disc is in steady state, 
so that the intrinsic emission from the outer disc (which 
forms the lower limit to the optical flux) is from the same 
mass accretion rate as is required to power the observed simultaneous 
X-ray emission.  Instead, the \textit{HST} data in T4 can be much 
better fitted by ignoring the simultaneous X-ray data, i.e., 
by assuming that the disc is not in steady state so that 
the mass accretion rate can vary with radius. 
The right panel of Figure~\ref{SED4-imp} shows 
how the curvature across the entire optical/UV bandpass 
can be modelled by very weak illumination of a disc with 
an intrinsically higher mass accretion rate than that 
consistent with the X-ray emission.  The curvature in the UV region 
is then almost entirely from the increasing colour temperature 
correction rather than from irradiation.  
The disc outer radius is again significantly smaller than 
in T1--3.  The estimated parameters in these additional 
two SED fittings are given in Table \ref{parameter2}.  
Thus there is a potential solution to explain the shape of
the \textit{HST} spectrum in T4 if the outer disc is 
intrinsically hotter than expected from the mass 
accretion rate as measured from the inner disc. 

\begin{figure*}
\begin{center}
\begin{minipage}{.49\textwidth}
\includegraphics[width=8.0cm]{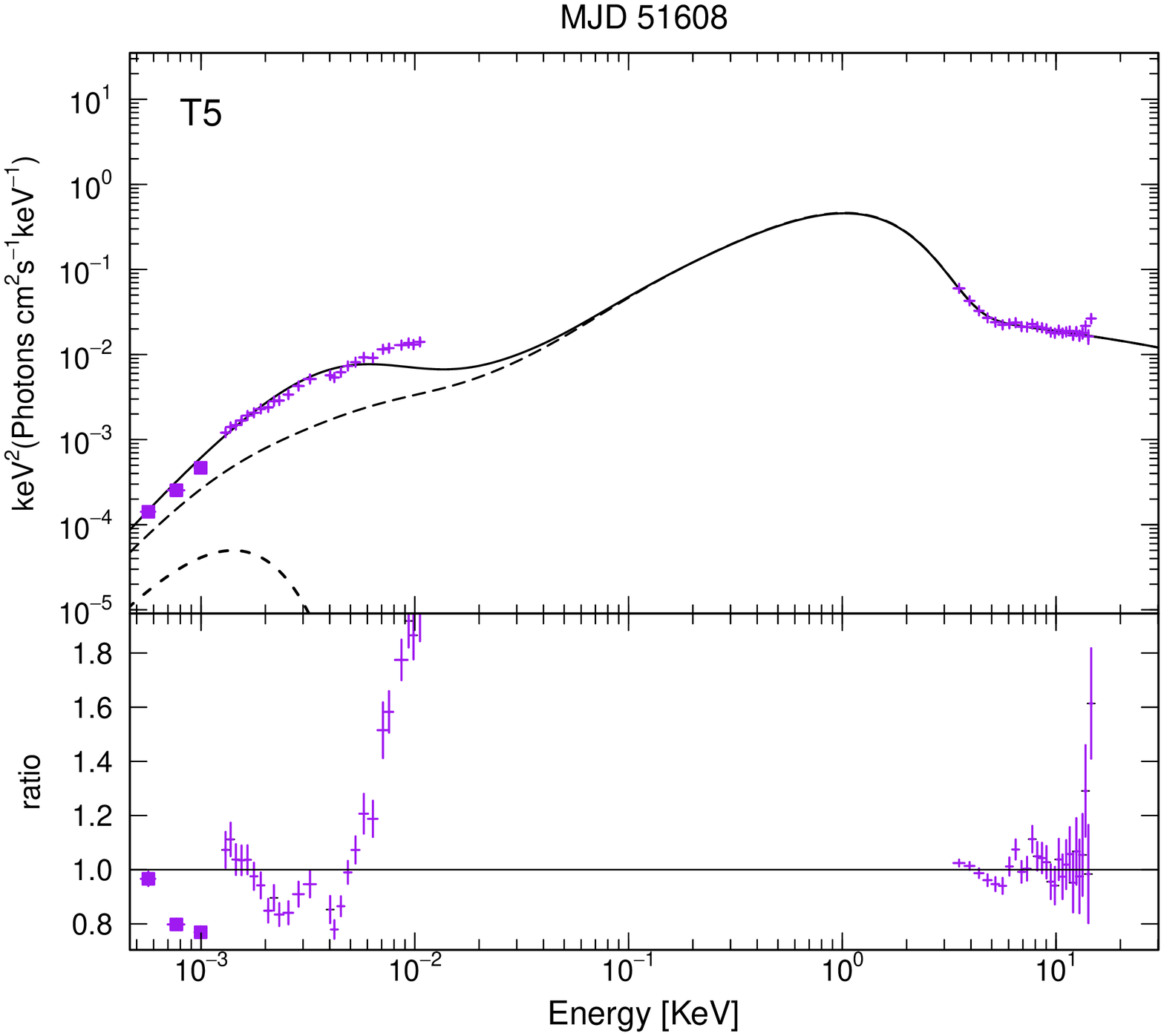}
\end{minipage}
\begin{minipage}{.49\textwidth}
\includegraphics[width=8.0cm]{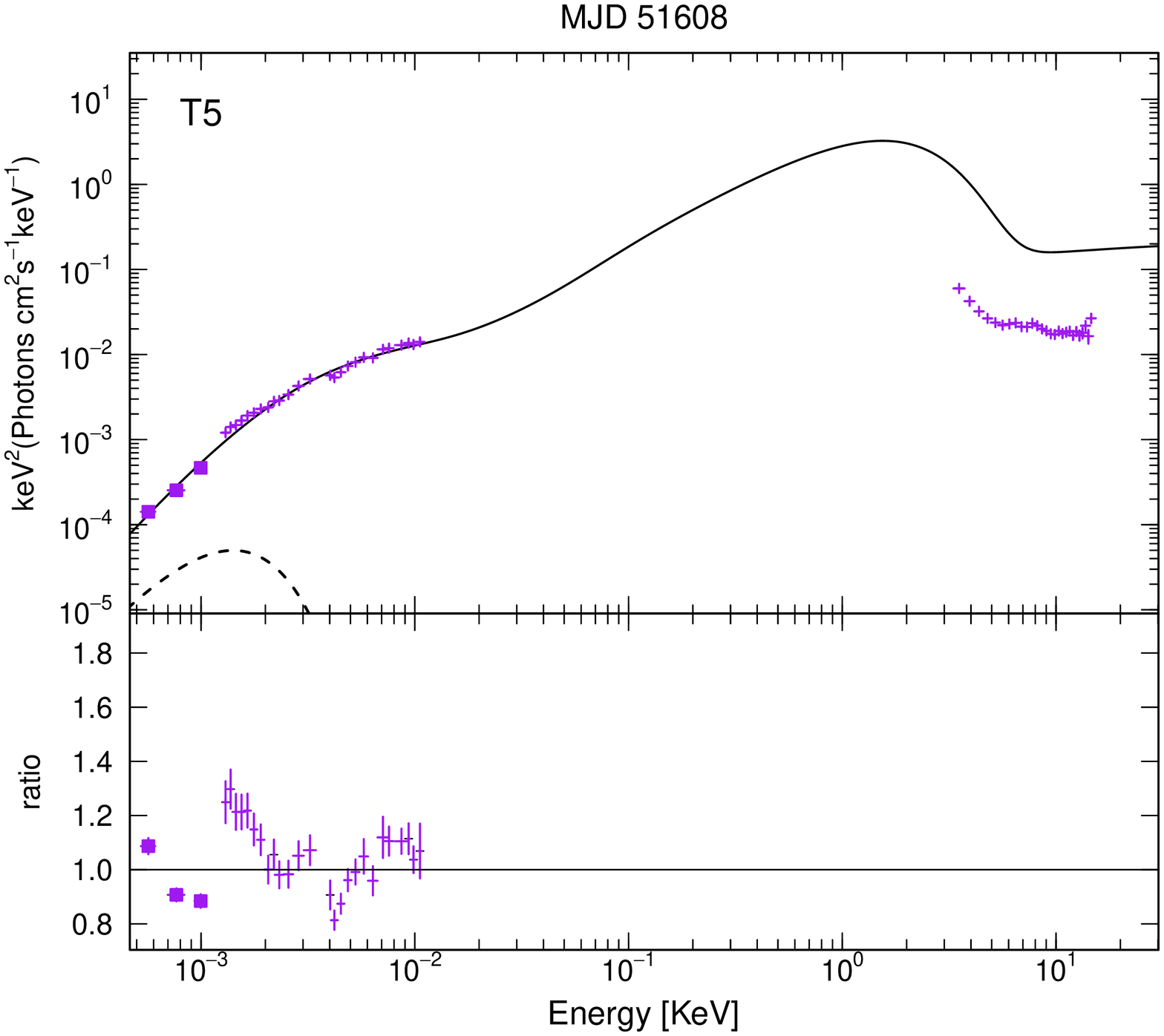}
\end{minipage}
\end{center}
\caption{Multi-wavelength SED modeling and the ratio of SED on T5 with fixed $f_{\rm out}$ and log($R_{\rm out}/R_{\rm g}$) (left panel) and those without the X-ray data (right panel).  The crosses and solid line represent the observed SED and the emission reproduced by our model, respectively.  The rectangles are the IR data.  The dashed line represents the emission from the companion star.  The long-dashed line in the left panel represents an unirradiated disc.  }
\label{SED5-imp}
\end{figure*}

The break in behaviour from T1--3 to T4 is even more marked in T5.  
The ratio R45 clearly has a strong energy dependence, 
with the UV flux changing by more than the optical flux 
(see the bottom panel of Figure \ref{fluxratio}).  
We again show the effect of fixed  $f_{\rm out} = 4.0 \times 
10^{-2}$, but this time
with fixed log($R_{\rm out}/R_g$) = 5.0 as seen in T4. 
This fits adequately (to within 20\%) to the optical and IR 
flux, but the UV excess becomes even larger than in the previous 
section (compare the bottom panel of Figure \ref{SED1} and 
the left panel of Figure \ref{SED5-imp}).  This is why 
the best-fit irradiated disc model in the previous section had
a larger $f_{\rm out}$ value. 

The right panel of Figure \ref{SED5-imp} shows the result of 
fitting to the UV/optical/IR data without including the X-ray 
emission, i.e., allowing the disc to be non-steady state.  
We get a much better fit to the curvature in the optical/UV 
spectra by an unirradiated, but hotter intrinsic outer disc, 
with $\log(R_{\rm out}/R_{\rm g}) \sim 5.0$ as well as T4.  Then, 
the best-fit values of $\log(L/L_{\rm Edd})$ and $f_{\rm out}$ 
are $-$1.09 and 1.1$\times$10$^{-8}$, respectively.  
However, the fit here is never statistically acceptable 
unlike that in T4.  The $\chi^2/$~dof is equal to 6.9, though 
the residuals are less than $\sim$20\%, and could be 
potentially produced by variability between different 
wavelengths in the \textit{HST} ranges.  X-ray spectral changes are most 
rapid around the transition \citep[see e.g.,][]{dun10BHBspectra}, 
which would mean that the reprocessed flux could also 
change rapidly.  Nonetheless, the optical/UV spectra are
better matched by the intrinsic disc emission than 
by reprocessed flux.  If this is intrinsic flux, 
the level implies a mass accretion rate through the outer 
disc is $\sim$5 times larger than that in the inner 
disc.

\begin{table}
	\centering
	\caption{Best-fit parameters of our additional SED modellings on T4.  See the notations in Table \ref{parameter} for the notation of each parameter.  }
	\label{parameter2}
	\begin{tabular}{cccc}
		\hline
		Model & Parameter & T4 ($f_{\rm out}$ fix) & T4 (only \textit{HST} spectra)\\
		\hline
optxrplir & log($L$/$L_{\rm Edd}$) & $-$1.67 & $-$0.80$\pm$0.08 \\
          & log($R_{\rm out}$/$R_{\rm g}$) & 5.01$\pm$0.01 & 5.00$\pm$0.03 \\
          & $f_{\rm out}$ & 4.0$\times$10$^{-2}$ (fix) & (1.8$\pm$1.4)$\times$10$^{-3}$ \\
\hline
\end{tabular}
\end{table}

\section{Discussion}

\subsection{Time Evolution of Irradiation Effect and Disc Size}

The UV/optical/IR spectra on T1--3 are mainly explained 
by the canonical irradiation to a steady state disc in our 
model (see also Sec.~6.1--6.2), although the IR emission 
requires an additional component, probably from the jet, 
in T1 and T2.
The amount of irradiation drops by 25\% as the source 
luminosity declines by a factor of 2 from T1 to T3.  
By contrast, our irradiation model fails to match 
the shape of the UV/optical spectra together with
the X-ray spectra in the late stage of the outburst 
(T4--5, see also Sec.~6.2--6.4).  
We firstly investigate theoretical ideas of how 
the irradiation should change, and then apply them 
to our data.

\subsubsection{Theoretical Models of Irradiation}

Here, we consider the structure of the irradiated disc 
to see if the observed $f_{\rm out}$ values can be 
matched by theoretical predictions. 
At first, we consider the irradiated disc model of 
\citet{cun76irradiation}.  This has a prediction for 
the scale height of the outer disc when irradiation 
dominates. 
Recasting their equation (27c) into the units used here, gives
\begin{equation*}
\begin{split}
H/R &= 1.5\times 10^{-3} (L/L_{\rm Edd})^{1/7} {(M_{\rm BH}/M_{\odot})}^{-1/7}
(R_{\rm out}/R_{\rm g})^{2/7} (R/R_{\rm out})^{2/7}\\
 &= f_{\rm d, out} (R/R_{\rm out})^{2/7}.  
\end{split}
\end{equation*}
This is derived by assuming $1/3$ of the flux 
thermalises in the disc, which is larger than 
the $(1-a)=0.1$ assumed here, but this will not affect
predictions of changes in $f_{\rm out}$.

Our source flux drops by a factor 2 from T1--3, so this 
equation predicts that this should cause a correlated 
decrease in $f_{\rm out}$ of around 10\%.  This is very 
close to that observed. 
However, the absolute values of $f_{\rm out}$ underestimate 
that seen in the data by around a factor 2. 
\textcolor{black}{ 
We note that the larger thermalising flux assumed in 
\citet{cun76irradiation} will slightly underestimate 
$f_{\rm out}$, so it suggests that there is an additional 
source of irradiation of the outer disc as well as direct 
illumination.  
\citet{beg83wind} show that X-ray irradiation of the disc 
can result in a corona and wind.  These features have some 
vertical scale height, so can scatter part of the central 
X-ray luminosity back down onto the outer disc.
We briefly outline the theory of these X-ray heated winds 
and coronae, and then make a quantitative estimate of 
the additional irradiating flux produced by scattering 
in these structures.  
}

\textcolor{black}{
Irradiation heats the disc surface to the Compton temperature,
$kT_{\rm IC}=\frac{1}{4} \int EL(E) dE/\int L(E)dE$, which 
depends only on the shape of the radiation field and not on 
its luminosity.  This forms a corona above the disc with scale
height set by the ratio of the sound speed $c^2_{\rm s,IC} = 
kT_{\rm IC}/\mu$ (where $\mu$ is the mass) to the escape 
velocity $v^2_{\rm esc} \sim GM/R$.  Wherever these are equal, 
the material can escape as a wind.  This defines the wind 
launch radius as $R_{\rm IC} = GM/{c_{\rm IC}}^2$.  
At smaller radii the material forms 
a static corona with scale height $H/R\sim c_{\rm s,IC}/v_{\rm esc} 
\sim (2R^3/R_{\rm IC})^{1/2}$ \citep{beg83wind}.
}

\textcolor{black}{
The wind region is a little more complex than the static corona 
as it is expanding so there is only finite time for heating.  
The heating rate depends on luminosity, so the detailed
properties of the wind then depend on luminosity as well as 
spectral shape.  
%The wind region will only be heated fast enough to reach 
%$T_{\rm IC}$ at $R_{\rm IC}$ if the luminosity is above 
%the critical luminosity $L_{\rm crit} \approx 0.03 {T_{\rm IC,8}}^{-1/2} 
%L_{\rm Edd}$ \cite[][where $T_{{\rm IC},8} = 
%T_{\rm IC}/10^8$ K]{don18wind,beg83wind}.  
This condition on the heating rate splits the wind region 
$R>R_{\rm IC}$ into three separate zones \citep{beg83wind,don18wind}. 
For $L>L_{\rm crit}$ the wind is heated to $T_{\rm IC}$ so is 
launched at $R_{\rm IC}$ (isothermal free wind, region A). 
At larger radii the heating rate decreases as the flux decreases, 
so eventually the wind is only heated to $T_{\rm ch}<T_{\rm IC}$.  
However, it still escapes as gravity is lower at these large radii 
(steadily heated wind, region B).
Instead, if $L<L_{\rm crit}$ then the material is heated only 
to $T_{\rm ch}<T_{\rm IC}$ at $R_{\rm IC}$ so it cannot escape
(gravity inhibited, region C) but at larger radii, the escape 
speed for gravity decreases faster than the decrease in 
characteristic temperature with flux so the wind can be 
launched (region B again). 
}

\textcolor{black}{
Hence the transition between the wind and atmosphere (i.e., 
wind launch radius) marks the boundary between region B and C, 
which is at $R \sim (L/L_{\rm crit})^{-1} R_{\rm IC}$ 
\cite[with some dependence on Compton temperature,][]{woo96coronae,hig15coronae,don18wind}.  
A more careful treatment shows that the material escapes 
as a wind when $R_{\rm in} > 0.2 R_{\rm IC}$ for 
$L > L_{\rm crit}$, and when $R_{\rm in} > 0.2 
(L/L_{\rm crit})^{-1} R_{\rm IC}$ for $L < L_{\rm crit}$. 
Hence there are only two regions where the material escapes 
efficiently as a wind, region A (where the wind is heated 
to $T_{\rm IC}$ as $L>L_{\rm crit}$)
and region B (where $T_{\rm ch} < T_{\rm IC}$ but $c_{\rm ch} > 
v_{\rm esc}$). 
}

\textcolor{black}{
In the discussion above, the wind and corona are smoothly 
connected. 
However, the disc has fairly low scale height, so the direct 
illumination is always at grazing incidence. This means that 
optical depth effects can become important, where the inner 
corona becomes optically thick to grazing incidence paths, 
and casts a shadow. This suppresses corona formation by 
direct illumination until the shape of the disc photosphere 
itself rises above the shadow \citep{beg83wind2}. 
Thus the inner corona and outer corona/wind form separate structures.
The inner corona does scatter some fraction of the central 
radiation, forming a diffuse source. 
However, this source is at small radii, where the scale height 
of the material is low due to the strong gravity, so this
diffuse source is not much larger than the intrinsic source 
size so does not contribute to increasing the illumination 
of the outer disc.   
}

On the other hand, scattering from the wind may form 
a larger scale height structure. 
The scattered flux from the wind forms a diffuse source 
of radiation which is maximised at the launch radius 
$R_{\rm in}$, where it has typical height $H_{\rm w}(R_{\rm in}) 
\approx R_{\rm in}$.  This forms an additional X-ray source 
which can efficiently illuminate the outer disc.  
\citet{beg83wind2} show that the scattered luminosity is 
\begin{equation*}
\frac{L_{\rm sc}({\rm wind})}{L}
= \frac{L}{L_{\rm Edd} \Xi_{\rm c,max}} 
\begin{cases}
0.5 & (A) \\
\Bigl( \frac{L_{\rm crit}}{L} \Bigr)^{2/3} 
[11-7(R_{\rm out}/R_{\rm IC})] & (B) \\
9 [1+\ln (R_{\rm out}/R_{\rm IC})] & (C) \\
\end{cases}
\end{equation*}
for wind regions A, B and C, respectively \citep[][their 
equation (3.9)]{beg83wind2}, where $\Xi_{\rm c,max}\approx 10$ 
is the maximum ionisation parameter on the cold branch. 
We note that this approximation is not piecewise 
continuous, nor does it take into account the increase 
in the wind mass loss expected as the source approaches 
$L_{\rm Edd}$ due to the significant radiation pressure.

Thus there are two components to the flux seen by the outer disc.  
One of them is direct illumination from the central source, 
which depends on the flaring of the disc and is $\propto 
H_{\rm d}/R$, and the other is from scattering in the outer 
corona/wind $\propto L_{\rm sc}({\rm wind}) (H_{\rm w}/R) 
\sim L_{\rm sc}({\rm wind})$. 
Thus the total flux seen by the outer disc can be approximated 
as $L+L_{\rm sc}({\rm wind})$ i.e. giving an effective $f_{\rm out} 
\approx f_{\rm d,out} + L_{\rm sc}({\rm wind})/L$.

\subsubsection{Application to XTE J1859$+$226}

We evaluate $T_{\rm IC}$ from the spectral shape for T1--5, 
using a maximum energy of 100~keV for the flux integration to 
mimic the effect of the Klein-Nisina reduction in Compton 
scattering cross-section. 
We use this to derive $L_{\rm crit}$ and $R_{\rm IC}$ 
for all of our datasets.  We tabulate the predicted 
thermal wind parameters in Table \ref{tab:thermal}. 
There is a marked drop in $T_{\rm IC}$ linked to an increase 
in $L_{\rm crit}$ and $R_{\rm IC}$ as the source softens from 
the very high state at T1--2 to the bright high/soft state at T3, 
but all these bright spectra (T1--3) have $L > L_{\rm crit}$, 
and hence have a wind region which starts at $R_{\rm in} = 
0.2R_{\rm IC}$.  Also, the wind region is A, since 
$R_{\rm out} > R_{\rm IC}$ at the outer disc radius.
%\footnote{Here, 
%$R_{\rm iso}$ is defined as $R_{\rm IC} (L/L_{\rm crit})$ 
%by equation (2.15) in \citet{beg83wind}.}  
The ignition radius of the wind is just within the outer 
region of the disc in these data, but the expected mass loss 
rate is fairly low as the outer disc radius is fairly small.  

The values for $f_{\rm out}$ from combining the theoretical 
models of scale height of the irradiated outer disc and 
scattering from the wind are remarkably close to those 
observed in T1--3.  
The absolute values are slightly dependent on the albedo assumed
(see the discussion in Sec.~7.1.1), but the predicted changes 
in $f_{\rm out}$ should be more robust. 
Importantly, the models predict similar decrease 
in $f_{\rm out}$ as seen in the data as the source flux and 
$T_{\rm ic,8}$ drop from T1 to T3.

Conversely, T4 has a much lower luminosity.  It is still 
disc dominated, but $T_{\rm IC}$ is low because of the cooler 
disc.  Also, $R_{\rm in} > R_{\rm out}$ because $L < 
L_{\rm crit}$.  
Thus there should be no wind region, only a static
corona heated to $T_{\rm IC}$.  The disc scale height should
be smaller intrinsically due to the lower luminosity, and 
there should be almost no additional scattered flux as there 
is very little wind. 
\textcolor{black}{
Yet the best fit to the data gives a much larger $f_{\rm out}$, 
since it (poorly) fits the observational UV flux under 
the assumption of a steady state disc, with the same mass 
accretion rate powering the intrinsic X-ray and optical/UV 
emission (see also Sec.~6.2 and 6.4).  
}

\textcolor{black}{
T5 has similarly low luminosity, though the X-ray tail is 
somewhat stronger (though quite poorly constrained), giving 
$T_{\rm IC}$ which is comparable to T3. However, the lower
luminosity again means that $L < L_{\rm crit}$. 
There is neither strong wind region nor additional source of 
illuminating flux to increase $f_{\rm out}$. 
Thus, like T4, the models predict that $f_{\rm out, d}$ is 
small due to the low luminosity, and 
$f_{\rm out} \sim f_{\rm out,d}$.  Yet again the data require 
a much larger $f_{\rm out}$ than seen in T1--3 in order to 
(very poorly) fit the observational UV flux under the assumption 
of steady state in fitting broadband data simultaneously 
(see also Sec.~6.3 and 6.4).  
}

\textcolor{black}{
Thus the irradiated disc theory, where the irradiation is 
a combination of direct illumination from the bright central 
regions, and its scattered flux from a wind, matches very well 
to the initial spectra in T1--3. 
However, these models predict that we should see a smaller 
reprocessed fraction in the dimmer spectra in T4--5, yet 
the data require a larger fraction to poorly (T4), or very 
poorly (T5) fit the data (see Sec.~6.2--6.4). 
While T5 may be more complex due to its proximity to the spectral 
transition, and constraints from this spectrum are less secure 
due to the non-simultaneity of the X-ray data, T4 is a classic 
disc dominated state, and the X-ray and UV/optical data were 
simultaneously observed.  Thus T4 would show robustly that
the irradiated steady state disc models are broken.  
As proposed in Sec.~6.4, the intrinsic heating of the outer 
disc is possibly larger than expected for the same mass accretion 
rate as seen in the inner regions.  
}

\begin{table}
	\centering
	\caption{Thermal corona/wind parameters in T1--5.  We use the luminosity and $R_{\rm out}$ summarised in Table \ref{parameter} in these calculations.  }
	\label{tab:thermal}
	\begin{tabular}{cccccc}
		\hline
Parameters & T1 & T2 & T3 & T4 & T5 \\
\hline
$T_{\rm IC}$$^{*}$ & 1.44 & 1.65 & 0.85 & 0.38 & 0.80 \\
$R_{\rm in} / R_{\rm g}$$^{\dagger}$ & 8.0$\times$10$^{4}$ & 7.0$\times$10$^{4}$ & 1.3$\times$10$^{5}$ & no wind & no wind \\
$L_{\rm est} / L_{\rm crit}$$^{\ddagger}$ & 5.7 & 3.7 & 2.2 & 0.15 & 0.13 \\
$\dot{M}_{\rm wind} / \dot{M}_{\rm acc}$$^{\S}$ & 0.29 & 0.31 & 0.09 & -- &  -- \\
%$f_{\rm out, cor}$$^{\P}$ & 0.041 & 0.038 & 0.030 & 0.014 & 0.015 \\
%$f_{\rm out, wind} / f_{\rm out, cor}$$^{|}$ & 0.12 & 0.08 & 0.08 & -- & -- \\
%$f_{\rm out}$$^{**}$ & 0.046 & 0.041 & 0.032 & 0.014 & 0.015 \\
$f_{\rm d, out}$$^{\P}$ & 0.032 & 0.029 & 0.028 & 0.017 & 0.016 \\
$R_{\rm ia}$$^{|}$ & 445 & 326 & 410 & 196 & 106 \\
%$L_{\rm sc, corona}/L_{\rm est}$ & 0.041 & 0.038 & 0.030 & 0.014 & 0.015 \\
$H_{\rm c}(R_{\rm ia})$$^{**}$ & 21 & 14 & 15 & 3.2 & 1.9 \\
%$R_{\rm in}$ = $H_{\rm wind} & 82770 & 73570 & 147147 \\ 
$L_{\rm sc, wind}/L_{\rm est}$$^{\dagger\dagger}$ & 0.021 & 0.013 & 0.010 & -- & -- \\
%$L_{\rm sc, wind}/L_{\rm est}$$^{**}$ & 0.10 & 0.079 & 0.12 & -- & -- \\
$f_{\rm out}$$^{\dagger\dagger}$ & 0.053 & 0.050 & 0.041 & 0.017 & 0.016 \\
\hline
\multicolumn{6}{l}{\parbox{240pt}{$^{*}$Inverse-Compton temperature estimated from the spectral shape.  In units of keV.  }}\\
\multicolumn{6}{l}{\parbox{240pt}{$^{\dagger}$Innermost radius where the wind is efficient.  }}\\
\multicolumn{6}{l}{\parbox{240pt}{$^{\ddagger}$Ratio of the estimated luminosity against the critical luminosity, with which the wind can be triggered.  }}\\
\multicolumn{6}{l}{\parbox{240pt}{$^{\S}$Ratio of the wind accretion against the whole accretion rate.  }}\\
\multicolumn{6}{l}{\parbox{240pt}{$^{\P}$Scale height of an outer disc defined by equation (27c) in \citet{cun76irradiation}.  }}\\
\multicolumn{6}{l}{\parbox{240pt}{$^{|}$Radius of an inner attenuation zone derived from equation (2.23) in \citet{beg83wind2}.  }}\\
\multicolumn{6}{l}{\parbox{240pt}{$^{**}$Typical height of an inner attenuation zone derived from equation (2.6) and $R_{\rm ia}$ in \citet{beg83wind2}.  }}\\
\multicolumn{6}{l}{\parbox{240pt}{$^{\dagger\dagger}$Scattered luminosity from a wind according to equation (3.9) in \citet{beg83wind2}.  }}\\
\multicolumn{6}{l}{\parbox{240pt}{$^{\ddagger\ddagger}$Predicted total $f_{\rm out}$ estimated by $f_{\rm d, out} + L_{\rm sc, wind}/L_{\rm est}$.  }}\\
\end{tabular}
\end{table}

\subsection{Time Evolution of Outer Disc Radius}

Another key difference between T1--3 and T4--5 is 
that the disc outer radius has clearly decreased 
by $\sim$15\% 
\citep[see also][]{hyn02j1859}.  Some disc
shrinkage is expected in the standard irradiated 
disc instability model \citep{dub01XNmodel}, but 
not as much as the factor 2 seen in these data.  
However, we can understand the degree of shrinkage 
by considering that disc extended much larger at 
the initial stage of the outburst than expected 
by the thermal instability.  
The thermal-tidal instability in low mass ratio 
objects makes it possible \citep{ich93SHmasstransferburst}.  

The thermal instability triggers a mild increase 
in disc size, but if this crosses the radius at which 
it goes into 3:1 resonance with the binary orbit then 
the disc is unstable against a non-axisymmetric perturbation.  
This is a slowly growing mode, which forms an eccentric disc 
which precesses in the orbiting frame as its period is 
slightly longer than the orbit.  This produces the superhump 
modulation seen in the optical light curves. The elliptical 
disc enhances the angular momentum transport with the companion 
star, which drives mass accretion through the outer disc, 
resulting in fast shrinkage of the disc.  
This combined thermal-tidal instability mechanism was 
first proposed for the SU UMa-type dwarf novae which 
typically have $q \lesssim 0.25$ and orbital periods of 
a few hours \citep{whi88tidal,osa89suuma}.
Even more extreme mass ratio systems such as the WZ Sge-type 
dwarf novae, an extreme subclass of SU UMa-type dwarf novae, 
with $q \lesssim 0.08$, can trigger a tidal instability 
at the 2:1 resonance radius.  The radius is larger than 
the 3:1 resonance radius, and this tidal instability grows 
faster.  It forms spiral 
arms with period very close to the orbital period, and 
early superhumps are believed to be its representation 
\citep{osa02wzsgehump}. 
Whichever tidal instability is triggered, the additional 
angular momentum transport and heating of the outer disc 
can keep it above the level of the hydrogen ionisation 
instability, so can give long, exponential decline light 
curves even without X-ray irradiation \citep{osa89suuma}. 

XTE J1859$+$226 has an extremely small mass ratio, 
$q=M_{2}/M_{\rm BH} \sim 0.09$, and its orbital period 
is 6.6 hours \citep{cor11j1859}.  In this system, 
the disc radius will definitely exceed the 3:1 resonance 
radius, $\sim$0.47$a$ in this object, so should trigger 
the thermal-tidal instability.  
Also, since the estimated disc radius in T1 exceeds 
the 2:1 resonance radius at $\sim$0.61$a$ in this object, 
so we should see early superhumps with period almost 
equal to the orbital period at the early stage of its 
outbursts, and ordinary superhumps later in the outburst 
with period slightly longer than the orbit.  
\textcolor{black}{Although 
there are no time-resolved optical light curves to 
unambiguously test this, \citet{uem04j1859} and \citet{zur02j1859} 
found some optical modulations with timescale close to 
the orbital period, which they suggested to be superhump 
candidates.  In addition, that outburst showed reflares 
according to \citet{zur02j1859}, 
which are one of characteristic features in the outbursts 
of WZ Sge-type stars \citep{kat15wzsge}.  }

\subsection{Irregular Disc Geometry at the Late Outburst}

\begin{figure}
\begin{center}
\includegraphics[width=8.0cm]{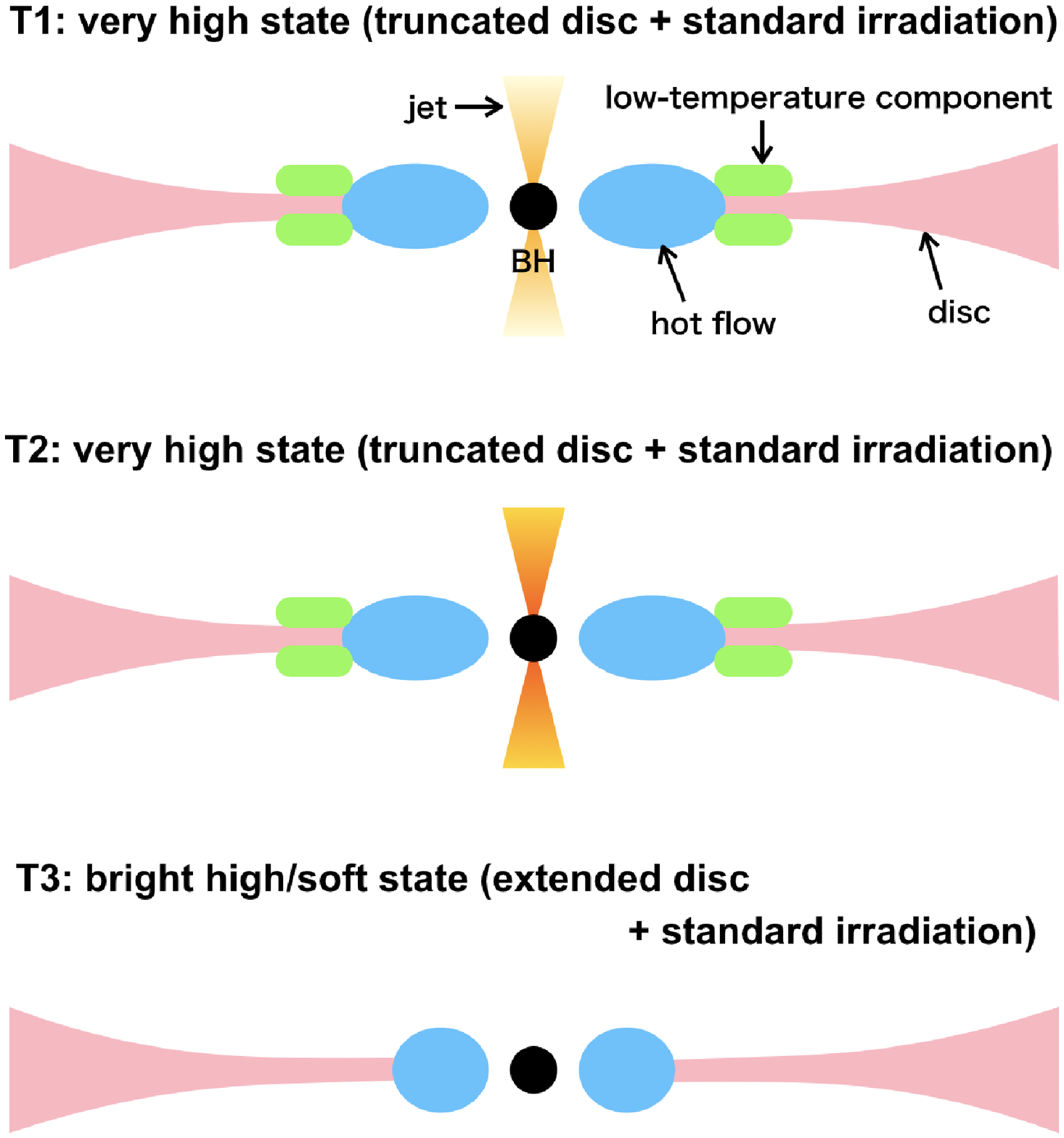}
\\
\includegraphics[width=8.0cm]{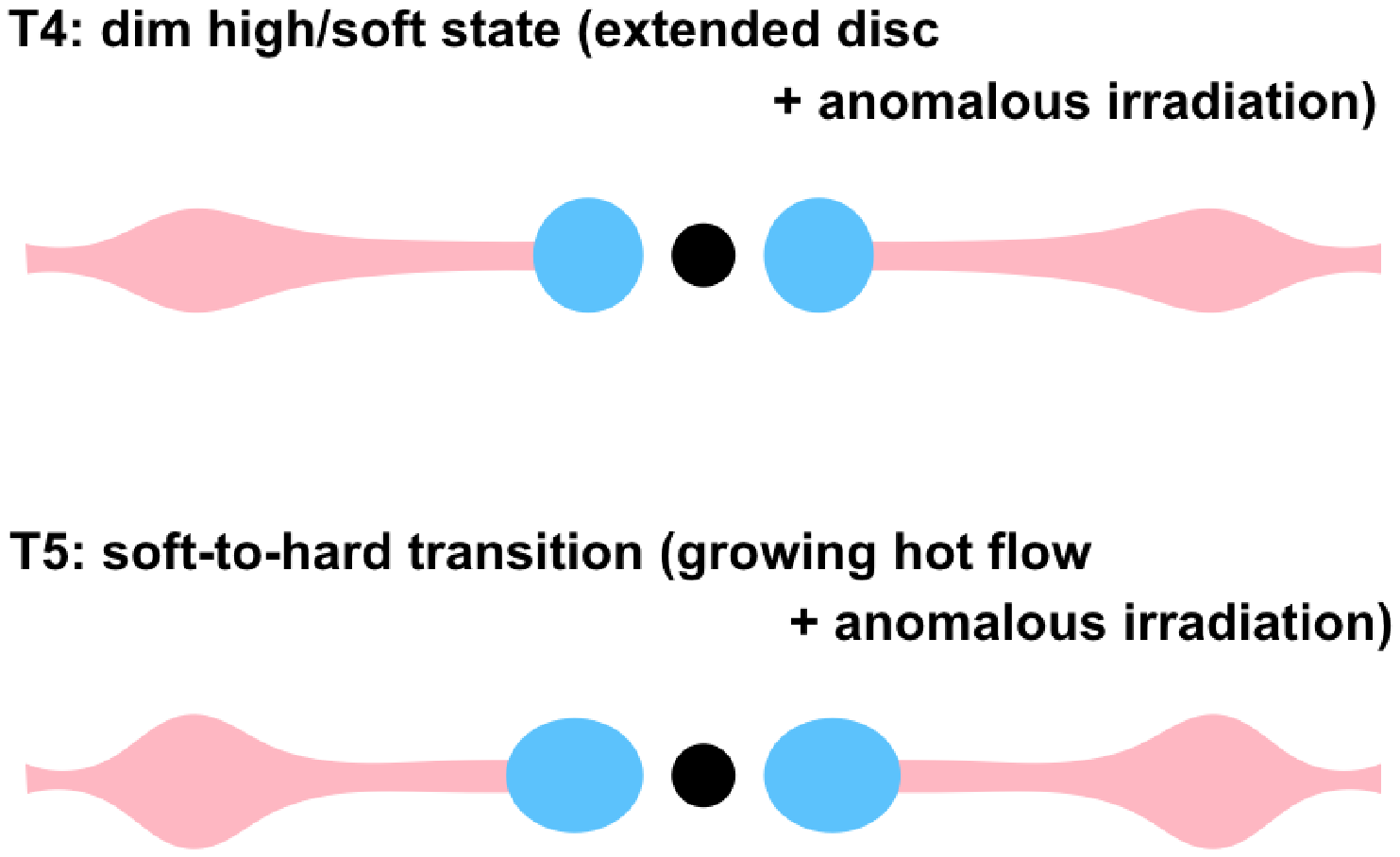}
\end{center}
\caption{Schematic picture of time-evolved accretion geometry in the 1999--2000 outburst of XTE J1859$+$226.  The black circle is a black hole.  The pink, blue, green, and orange structures represent an accretion disc, an optically-thin hot flow, a low-temperature optically-thick Comptonization component, and jet ejection, respectively (colour online).  }
\label{geometry}
\end{figure}

On T4 and T5, we do not obtain reasonable fits, and 
the observations are inconsistent with the theoretical 
predictions as described in Sec.~7.1.2.  
Considering that other emission mechanisms like jet ejections 
are unlikely, we suggest that the assumptions in our model 
are not applicable for T4 and T5, i.e., irregular outer disc 
geometry possibly producing non-steady state would be formed 
in this outburst.  
%As explained in the previous section, in dwarf novae, 
%the thermal-tidal instability naturally explains the slow decline 
%of optical light curves without the effect of X-ray irradiation. 
%Instead, the enhanced angular momentum transport by tidal forces 
%exerted by the secondary star at the outer disc leads to enhanced 
%heating, keeping hydrogen ionised and suppressing the cooling wave 
%which would otherwise switch the disc back into quiescence.  
Here we consider the thermal-tidal instability possibly forms 
an anomalous outer disc geometry.  
To date, the only calculations of the thermal-tidal instability 
in black hole binaries did not include the interplay of this 
with the X-ray irradiation heating \citep{ich94bxhn}, but they 
showed that the disc has a complex temperature dependence 
(see Figures 3 \& 5 in that paper).  
In addition, the tidal heating could increase the scale height 
of the outer disc, and/or changes its shape from the expected 
$H/R\propto R^{2/7}$ due to the piled-up mass at the outer 
disc \citep{osa89suuma}.  
Thus the effect can produce weird outer disc geometry 
related to non-steady state that we propose at T4--5 as sketched 
in Figure \ref{geometry}, 
although the model in \citet{ich94bxhn} predicts a rather complex 
temperature dependence in the outer disc, whereas the spectra in 
T4 and T5 are best fit by the standard temperature distribution 
in the outer disc, with just an increased mass accretion rate 
relative to the inner disc.  
In the future, full calculations should include the effect of 
X-ray heating and see whether this in combination with 
the thermal-tidal instability can reproduce the observed data. 

There are not many other possibilities which can explain the hot 
outer disc as seen in T4 and T5.  
Wind losses should also be included since these necessarily 
reduce the mass accretion rate between the inner and outer disc.
However, the thermal wind is not strong in this system 
as the disc is small, and its effect should be biggest in T1--3, 
whereas the data most require a non-steady state disc in T4--5.  
\textcolor{black}{
The effect of magnetic winds is difficult to quantify, but there 
may not be a strong requirement for these in the black-hole 
binaries \citep{don18wind}.  }
\citet{hyn02j1859} suggest that the difference in optical/UV 
spectral shape arises from warping of the disc due to irradiation 
\citep{wij99warpdisk,ogi01XBwarpeddisk}; however, this seems 
unlikely because this kind of instability is most effective
in long-period (more than 1 day) systems.  
Full calculations are required in order to study the interplay of 
all these effects. 

These possibilities raised above may be testable with fast 
photometric data simultaneous with X-ray observations.
If the model in \citet{ich94bxhn} is correct, the optical/UV 
emission is dominated by intrinsic luminosity of the underlying 
disc at the outer region, and hence, it will vary only very slowly, 
and non-correlated optical variability inherent to the outer disc 
may be observable.  
On the contrary, if the emission is dominated by reprocessing, 
the optical/UV variations will significantly correlate with 
the X-ray fast variations. 
If the warp precessions are dominant, strong optical periodicity 
as well as fast optical variability from reprocessing would be 
observed.  
\textcolor{black}{If we solve this problem, we can know correctly 
the reason of the slow optical decline in outburst in LMXBs.  
Not only XTE J1859$+$226 but also many black-hole X-ray binaries 
show more rapid X-ray outburst decay than the optical decay 
\citep{che97BHXN}, so this is a common observational 
feature.
X-ray irradiation is considered to form the slow decay 
of optical outburst light curves in LMXBs \citep{dub01XNmodel}; 
however, if the irradiation effect is weak at the late stage of 
outbursts, the tidal forces by the companion stars would be 
plausible.  
In many systems, it has not yet been constrained well which of 
the two is dominant.  
}

\section{Conclusions}

We analyzed the 5 sets of broadband spectra from 
XTE J1859$+$226 during its 1999--2000 outburst 
by using the best current model of the irradiated disc 
continuum with the expected irradiation dominated
$H\propto R^{9/7}$ shape, giving $T(R) \propto R^{-3/14}$. 
This works extremely well in the bright stages of the outburst,
and is able to fit the X-ray, UV and optical continua
spectra, though the IR requires a small contribution 
from the jet whose power tracks the hard X-ray continuum.  
The solid angle subtended by the irradiated outer disc 
is observed to decrease by $\sim$25\% as the source flux
drops by a factor 2, and the spectrum softens from 
the very high state to the disc dominated high soft state. 
The changing solid angle implied our fits are exactly 
consistent with the predictions of scale height of 
the irradiated outer disc and scattering in 
a thermal wind at the early stage of the outburst, 
with an albedo of $\sim 0.9$.
%, as expected for 
%such strongly ionised material \citep[see 
%e.g.,][]{don18wind} and as inferred for other LMXBs 
%\citep{dej96reprocess}. 

The last two spectra, which are much lower in X-ray 
flux, are not well explained by these models.  
Since the optical and UV spectra have a different shape than 
predicted by the standard illuminated disc, it is difficult 
to precisely evaluate the amount of irradiation in these spectra.
Nonetheless, the straightforward model 
fits require that a larger fraction of the flux is intercepted 
by the outer disc. 
However, the irradiated disc models predict the opposite: the 
scale height of the outer disc becomes smaller at these lower 
luminosities, and the thermal wind models
predict that the wind should shut off at these low luminosities 
and temperatures. 
These late spectra are instead better fit by the unilluminated 
standard disc, which has $T\propto R^{-3/4}$.   
However, the models without illumination have 
an implied mass accretion rate through the outer disc, 
which is much higher than that which powers the inner disc.  
Our analyses also imply that the outer disc shrinks 
by more than a factor of 2. 

The observed large shrinkage of the disc can be triggered 
by enhanced dissipation due to the thermal-tidal 
instability.  
Since this system likely has an extremely small mass 
ratio, this kind of instability should work.  
The tidal instability may also explain the different 
optical/UV spectra seen as the outburst dims as it 
produces additional heating of the outer disc and 
the resulting temperature distribution can be far from 
the steady state $R^{-3/4}$.  However, there are 
no simulations to date of the thermal-tidal instability 
including the effects of X-ray illumination, so these
models are not self consistent. The tidal instability 
may also change the vertical structure of the outer disc 
in a non axisymmetric way, so that the disc geometry can be 
changed from that predicted by the irradiated disc models. 

Better models including all these effects 
during the outburst may be able to match the observed behaviour, 
while better data -- specifically including high time 
resolution UV photometry such as UVSat \citep{UVSat} -- 
should determine whether the UV flux in these dimmer spectra 
is due to X-ray irradiation or intrinsic heating of the outer disc.

\section*{Acknowledgements}

This work was financially supported by the Grant-in-Aid 
for JSPS Fellows for young researchers (M.~Kimura).  
C.~Done acknowledges the Science and Technology Facilities 
Council (STFC) through grant ST/P000541/1 for support.

%%%%%%%%%%%%%%%%%%%%%%%%%%%%%%%%%%%%%%%%%%%%%%%%%%



%\bibliographystyle{mnras}

%\bibliography{mn-jour,/Users/kimuchan/cvs}
%\bibliographystyle{mnras}

\newcommand{\noop}[1]{}

%%%%%%%%%%%%%%%%%%%%%%%%%%%%%%%%%%%%%%%%%%%%%%%%%%

%%%%%%%%%%%%%%%%% APPENDICES %%%%%%%%%%%%%%%%%%%%%

\appendix

%%%%%%%%%%%%%%%%%%%%%%%%%%%%%%%%%%%%%%%%%%%%%%%%%%

% Don't change these lines
\bsp	% typesetting comment
\label{lastpage}
\end{document}